\documentclass[a4paper,11pt]{article}
\pdfoutput=1
\usepackage{jheppub}
\usepackage[T1]{fontenc}
\usepackage{graphicx}
\usepackage{amsmath}
\usepackage{xspace}
\usepackage{subfig}
\usepackage{xcolor}
\usepackage{mathtools}
\usepackage{braket}
\usepackage{multirow}
\usepackage{colortbl}
\usepackage{setspace}
\usepackage{slashed}
\usepackage{appendix}
\usepackage{booktabs}
\usepackage{cancel}
\usepackage{array}
\usepackage{pst-node} 
\usepackage{tikz}
\usepackage{wasysym}
\usepackage{amssymb}
\usepackage{cleveref}
\usepackage{makecell}

\usepackage[normalem]{ulem}

\usepackage{float}
\usetikzlibrary{tikzmark}
\usetikzlibrary{positioning}

\allowdisplaybreaks

\usepackage[detect-all]{siunitx}
\newcommand{\ovbb}[0]{$0\nu\beta\beta$ }
\newcommand{\lr}[1]{\left( #1\right)}
\newcommand{\lrs}[1]{\big( #1\big)}
\newcommand{\hc}{\text{h.c.}}

\Crefname{table}{}{}
\Crefname{figure}{}{}
\definecolor{newgray}{gray}{0.85}
\definecolor{newgray2}{gray}{0.95}
\newcolumntype{P}[1]{>{\centering\arraybackslash}p{#1}} % for horizontal alignment with limited column width
\newcolumntype{M}[1]{>{\centering\arraybackslash}m{#1}} % for horizontal and vertical alignment with limited column width
\newcolumntype{L}[1]{>{\raggedright\arraybackslash}m{#1}} % for vertical alignment left with limited column width
\newcolumntype{R}[1]{>{\raggedleft\arraybackslash}m{#1}} % for vertical alignment right with limited column width
\newcolumntype{?}{!{\vrule width 0.8pt}}

\title{Radiative neutrino masses from dim-7 SMEFT:\\ a simplified multi-scale approach 
}

\author[a,b]{K\r{a}re Fridell}
\author[c,d,e,f]{Luk\'a\v{s} Gr\'af}
\author[g]{Julia Harz} 
\author[h]{Chandan Hati} 
\emailAdd{karef@post.kek.jp}
\emailAdd{lukas.graf@nikhef.nl}
\emailAdd{julia.harz@uni-mainz.de}
\emailAdd{chandan@ific.uv.es}
\affiliation[a]{\AddrKEK}
\affiliation[b]{\AddrFSU}
\affiliation[c]{\AddrNikhef}
\affiliation[d]{\AddrCUni}
\affiliation[e]{\AddrUCB}
\affiliation[f]{\AddrUCSD}
\affiliation[g]{\AddrJGU}
\affiliation[h]{\AddrIFIC}

\preprint{KEK-TH-2673, 
MITP-24-090}

\newcommand{\AddrKEK}{Theory Center, Institute of Particle and Nuclear Studies,\\ High Energy Accelerator Research Organization (KEK), Tsukuba 305-0801, Japan}
\newcommand{\AddrNikhef}{Nikhef, Theory Group, Science Park 105, 1098 XG Amsterdam, The Netherlands}
\newcommand{\AddrCUni}{Institute of Particle and Nuclear Physics, Faculty of Mathematics and Physics, Charles University in Prague, V Hole\v{s}ovi\v{c}k\'ach 2, 180 00 Praha 8, Czech Republic}
\newcommand{\AddrUCB}{Department of Physics, University of California, Berkeley, CA 94720, USA}
\newcommand{\AddrUCSD}{Department of Physics, University of California, San Diego, CA 92093, USA}
\newcommand{\AddrJGU}{$PRISMA^+$ Cluster of Excellence \& Mainz Institute for Theoretical Physics, FB 08 - Physics, Mathematics and Computer Science, Johannes Gutenberg-Universit\"{a}t Mainz, 55099 Mainz, Germany}

\newcommand{\AddrIFIC}{Instituto de F\'isica Corpuscular (IFIC), Universitat de Valencia-CSIC, E-46980 Valencia, Spain}
\newcommand{\AddrFSU}{Department of Physics, Florida State University, Tallahassee, FL 32306, USA}

\abstract{Lepton-number-violating interactions occur in the Standard Model Effective Field Theory (SMEFT) at odd dimensions starting from the dimension-5 Weinberg operator. Although the operators at dimension-7 and higher are more suppressed by the heavy new scale, they can be crucial when traditional seesaw mechanisms leading to tree-level dimension-5 contributions are absent. We identify all minimal tree-level UV-completions for dimension-7 $\Delta L=2$ SMEFT operators without covariant derivatives and propose a new simplified approach for estimating the radiative neutrino masses arising from such operators. This dimensional-regularisation-based approach provides a more accurate estimate for the loop neutrino masses when the new physics fields are hierarchical in mass, as compared to the cut-off-regularisation-based approach often employed in the literature. This allows us to identify viable regions of parameter space in the full list of relevant simplified models close to the current limits set by neutrinoless double beta decay and the LHC that would previously have been thought to be excluded by neutrino-mass constraints.
}
\begin{document}	
\maketitle
\flushbottom

%%%%%%%%%%%%%%%%%%%%%%%%%%%%%%%%%%%%%%%%%%%%%%%%%%%%%%%%%%%%%%%%%%%%%%%%%%%%%%%%%
\section{Introduction}
%%%%%%%%%%%%%%%%%%%%%%%%%%%%%%%%%%%%%%%%%%%%%%%%%%%%%%%%%%%%%%%%%%%%%%%%%%%%%%%%%
The observation of non-zero neutrino masses represents a key laboratory evidence for beyond-the-Standard-Model (BSM) physics. Possible theoretical explanations can be used as guidance in searches for new, UV degrees of freedom. If neutrinos are Majorana particles, the corresponding New Physics (NP) model inevitably needs to include interactions that do not preserve the lepton number. Therefore, observables testing this seemingly accidental global symmetry of the Standard Model (SM) provide a well-motivated probe of BSM scenarios.

At low energies, the NP effects can be encoded in terms of the effective operators invariant under the SM gauge symmetry, forming the Standard Model Effective Field Theory (SMEFT). Lepton number violation (LNV) occurs in this formalism at odd mass dimensions~\cite{Kobach:2016ami}. Therefore, the leading contribution is expected to stem from dimension-5 SMEFT consisting of the well-known Weinberg operator. The set of next-to-leading LNV interactions appear at dimension 7, the basis of which has been established not so long ago~\cite{Henning:2015alf,Lehman:2015coa,Henning:2015daa,Lehman:2015via,Fuentes-Martin:2016uol,Fonseca:2019yya}.

In Ref.~\cite{Fridell:2023rtr} we have discussed the phenomenological limits on dimension-7 $\Delta L = 2$ operators. These operators are constrained by a wide range of different searches, including e.g.\ the LHC, \ovbb decay, and rare kaon decays \footnote{See Ref.~\cite{Fridell:2023rtr} for an overview and also Refs.~\cite{Scholer:2023bnn,Bolton:2021pey,Bolton:2020xsm,Li:2020lba,Bolton:2019wta,Bell:2006wi,Liao:2019gex,Liao:2021qfj,Brdar:2020quo,Magill:2018jla,Berryman:2016slh} for some dedicated discussion of some intersting observables.} However, as pointed out therein, the effective approach is subject to a number of assumptions that may result in limited validity and possibly less realistic constraints in comparison with specific UV-completions. To counter these issues, we continue our investigation of dimension-7 $\Delta L = 2$ operators in this work by analysing the complete set of simplified models with up to two new fields that UV-complete these non-renormalisable interactions at tree-level (see also Ref.~\cite{Herrero-Garcia:2019czj}).

Despite their stronger suppression compared to the dimension-5 (Weinberg) operator, LNV at dimension 7 in SMEFT can also lead to sizable contributions for specific lepton-number-violating processes. Clearly, if the usual three seesaw fields leading to type-I, type-II, or type-III seesaw mechanisms are absent at high energies, the Weinberg operator is not generated at tree-level at low energies. In such a scenario, the leading contribution to LNV may stem from dimension-7 operators, which is a possibility previously proposed and discussed in a variety of publications~\cite{Bonnet:2009ej,Angel:2012ug,Cai:2014kra,deBlas:2017xtg,Cepedello:2017lyo,Cepedello:2017eqf,Anamiati:2018cuq,DeGouvea:2019wnq,Gargalionis:2020xvt,Chen:2021rcv,Herrero-Garcia:2019czj,Banerjee:2020jun,Chala:2021juk,DasBakshi:2021xbl}. Observations suggesting a higher-dimensional origin of LNV could hint at radiative neutrino Majorana mass models (see e.g.\ Refs.~\cite{Cai:2017jrq,Babu:2019mfe}).

When determining the neutrino mass contributions within the EFT framework, typically the cut-off-regularisation-based estimates are often employed in the literature, see e.g.~Ref.~\cite{deGouvea:2007qla} for some relevant discussion on the methodology. It can be easily seen from a direct comparison with neutrino mass expressions obtained in complete UV models that these simple estimates suffer from a number of caveats stemming from the assumptions necessary for application of the naive cut-off approximation. The most apparent discrepancy occurs in the case of a large hierarchy between the masses of the new fields. In this work, we address these drawbacks and investigate the loop neutrino masses from the underlying UV physics of the dimension-7 SMEFT. We propose a simplified approach based on dimensional regularisation and matching simple UV-completions to an intermediate EFT. This helps in getting a much closer approximation of the exact model-based results in contrast to the simple cut-off regularisation based estimates, when the heavy NP is hierarchical in mass. Allowing for the masses of the UV fields to be independent parameters opens up new parts of parameter space to be explored, which we use to derive neutrino-mass constraints in the simplified multi-scale approach setup and confront them with bounds from other relevant observables.

The paper is organised as follows: we review the EFT framework and list all the possible two-field UV-completions using the systematic approach of covariant derivative expansion (see Appendix~\ref{sec:CDE} for details of the calculation) in Section~\ref{sec:eft}. Next, we cover the neutrino mass generation mechanisms in all the listed UV-completions of the dimension-7 operators in Section~\ref{sec:EFTvsUVmass}, and provide an explicit comparison of neutrino masses obtained as a single-scale cut-off estimate with a full in-depth analysis of the SM extension with two scalar leptoquarks, $\tilde{R}_2$ and $S_1$ in Section~\ref{sec:leptoquark}. Following that, we introduce and discuss a multiscale EFT approach to radiative neutrino masses in Section~\ref{sec:dim_reg}. Eventually, we combine, compare, and analyse the constraints imposed on the studied simplified models by neutrino masses as well as other available observables in Section~\ref{sec:combUVLNV}, before we conclude in Section~\ref{sec:conclusions}.

%%%%%%%%%%%%%%%%%%%%%%%%%%%%%%%%%%%%%%%%%%%%%%%%%%%%%%%%%%%%%%%%%%%%%%%%%%%%%%%%%
\section{$\Delta L = 2$ dimension-7 SMEFT and tree-level UV-completions}
\label{sec:eft}
%%%%%%%%%%%%%%%%%%%%%%%%%%%%%%%%%%%%%%%%%%%%%%%%%%%%%%%%%%%%%%%%%%%%%%%%%%%%%%
With NP expected to be present at energy scales higher than that of electroweak symmetry breaking, it has become customary to view and treat the SM as an effective description of a UV-complete theory. In this picture, the low-energy SM Lagrangian can be extended by a corresponding effective part as 
\begin{equation}
	\label{eq:wilson}
	\mathcal{L}=\mathcal{L}_{\text{SM}} + \mathcal{L}_\text{eff}\, ,
\end{equation}
where
\begin{equation}
\label{eq:EFTlag}
    \mathcal{L}_\text{eff} = \sum_{d}\sum_a C_a^{(d)}\mathcal{O}_a^{(d)},
\end{equation}
with $C_a^{(d)}$ denoting the Wilson coefficients of the respective effective operators $\mathcal{O}_a^{(d)}$ of mass dimension $d$. Operators that contribute to Majorana neutrino masses must violate lepton number $L$ by two units, $\Delta L = 2$, and they appear only at odd mass dimensions. Therefore, the Lagrangian we consider in this work reads
\begin{equation}
	\label{eq:wilsonLNV}
	\mathcal{L}_\text{eff}=C^{(5)}\mathcal{O}^{(5)} + \sum_a C_a^{(7)}\mathcal{O}_a^{(7)} + \sum_a C_a^{(9)}\mathcal{O}_a^{(9)} + \dots \, .
\end{equation}
We focus specifically on dimension-7 operators, following the notation used in Ref.~\cite{Fridell:2023rtr}. For clarity, we show explicitly the definition of the whole $\Delta L = 2$ dimension-7 SMEFT basis in Tab.~\ref{tab:dim7op}.

\setlength{\extrarowheight}{4pt}
\begin{table}[t]
    \centering
    \begin{tabular}{|>{\raggedright}m{50pt}|>{\raggedright}m{50pt}|>{\centering}m{170pt}|c}
    \cline{1-3}
    Type & $\mathcal{O}$ & Operator & \\[2pt]
    \cline{1-3}
    $\Psi^2 H^4$ & $\mathcal{O}_{LH}^{pr}$ & $\epsilon_{ij}\epsilon_{mn}\lrs{\overline{L_p^c}{}^iL_r^m}H^jH^n\lrs{H^\dagger H}$ & \\[2pt]
     \cline{1-3}
    $\Psi^2H^3D$ & $\mathcal{O}_{LeHD}^{pr}$ & $\epsilon_{ij}\epsilon_{mn}\lrs{\overline{L_p^c}{}^i\gamma_\mu e_r}H^j\lrs{H^miD^\mu H^n}$ & \\[2pt]
     \cline{1-3}
    \multirow{2}{*}{$\Psi^2H^2D^2$} & $\mathcal{O}_{LHD1}^{pr} $ & $\epsilon_{ij}\epsilon_{mn}\lrs{\overline{L^c_p}{}^iD_\mu L^j_r}\lrs{H^mD^\mu H^n}$ & \\[2pt]
     \cline{2-3}
     & $\mathcal{O}_{LHD2}^{pr}$ & $\epsilon_{im}\epsilon_{jn}\lrs{\overline{L^c_p}{}^iD_\mu L_r^j}\lrs{H^mD^\mu H^n}$ & \\[2pt]
      \cline{1-3}
    \multirow{2}{*}{$\Psi^2H^2X$} & $\mathcal{O}_{LHB}^{pr} $ & $g\epsilon_{ij}\epsilon_{mn}\lrs{\overline{L_p^c}{}^i\sigma_{\mu\nu}L_r^m}H^jH^nB^{\mu\nu}$ & \\[2pt]
     \cline{2-3}
     & $\mathcal{O}_{LHW}^{pr} $ & $g'\epsilon_{ij}\lrs{\epsilon \tau^I}_{mn}\lrs{\overline{L_p^c}{}^i\sigma_{\mu\nu}L_r^m}H^jH^nW^{I\mu\nu}$ & \\[2pt]
     \cline{1-3}
    $\Psi^4D$ &  $\mathcal{O}_{\bar{d}uLLD}^{prst}$ & $\epsilon_{ij}\lrs{\overline{d_p}\gamma_\mu u_r}\lrs{\overline{L_s^c}{}^iiD^\mu L_t^j}$ & \\[2pt]
     \cline{1-3}
    \multirow{5}{*}{$\Psi^4H$} & $\mathcal{O}_{\bar{e}LLLH}^{prst}$ & $\epsilon_{ij}\epsilon_{mn}\lrs{\overline{e_p}L_r^i}\lrs{\overline{L_s^c}{}^j L_t^m}H^n$ & \\[2pt]   
     \cline{2-3}
    & $\mathcal{O}_{\bar{d}LueH}^{prst}$ & $\epsilon_{ij}\lrs{\overline{d_p}L_r^i}\lrs{\overline{u_s^c}e_t}H^j$ & \\[2pt]
     \cline{2-3}
     & $\mathcal{O}_{\bar{d}LQLH1}^{prst}$ & $\epsilon_{ij}\epsilon_{mn}\lrs{\overline{d_p}L_r^i}\lrs{\overline{Q_s^c}{}^jL_t^m}H^n$ & \\[2pt]
     \cline{2-3}
    & $\mathcal{O}_{\bar{d}LQLH2}^{prst}$ & $\epsilon_{im}\epsilon_{jn}\lrs{\overline{d_p}L_r^i}\lrs{\overline{Q_s^c}{}^jL_t^m}H^n$ & \\[2pt]
     \cline{2-3}
     & $\mathcal{O}_{\bar{Q}uLLH}^{prst}$ & $\epsilon_{ij}\lrs{\overline{Q_p}u_r}\lrs{\overline{L_s^c}L_t^i}H^j$ & \\[2pt]
     \cline{1-3}
    \end{tabular}
    \caption{List of all independent $\Delta L= 2$ operators at dimension-7 SMEFT extracted from the basis of Ref.~\cite{Lehman:2014jma,Liao:2016hru}. Here, $D_\mu x^n$ denotes $(D_\mu x)^n$ for $x \in \{L_i,H\}$.}
    \label{tab:dim7op}
\end{table}

%%%%%%%%%%%%%%%%%%%%%%%%%%%%%%%%%%%%%%%%%%%%%%%%%%%%%
%%%%%%%%%%%%%%%%%%%%%%%%%%%%%%%%%%%%%%%%%%%%%%%%%%%%%%%%%%%%%%%%%%%%%%%%%%%%%%%%%

We can expand the operators in Tab.~\ref{tab:dim7op} following the procedure of Ref.~\cite{Gargalionis:2020xvt}, which we extend by including heavy vector fields. Using the covariant derivative expansion (CDE) formalism~\cite{Henning:2014wua}, we identify the dimension-7 LNV operators in Tab.~\ref{tab:dim7op} with the terms found by systematically integrating out heavy BSM fields from a renormalisable Lagrangian. This lets us find all possible tree-level UV completions of the dimension-7 LNV operators. In Sec.~\ref{sec:combUVLNV} we use this method to expand upon the results of previous works~\cite{Bonnet:2009ej,Angel:2012ug,Cai:2014kra,deBlas:2017xtg,Cepedello:2017lyo,Cepedello:2017eqf,Anamiati:2018cuq,DeGouvea:2019wnq,Gargalionis:2020xvt,Chen:2021rcv,Herrero-Garcia:2019czj,Banerjee:2020jun,Chala:2021juk,DasBakshi:2021xbl} by systematically categorizing the UV-completions in terms of their corresponding operators, evaluating the neutrino masses for each model, and comparing these results with a wide set of phenomenological probes for a varying hierarchy in the internal degrees of freedom.

We start with a general Lagrangian $\mathcal{L}_\text{S}$ describing a simplified model extension to the SM, 
\begin{equation}
	\label{eq:lhLag}
	\mathcal{L}_\text{S} = \sum_{i}\lr{\mathcal{L}_{\pi_i}^{\text{kin}}+\mathcal{L}_{\pi_i}^{\text{int}}+\mathcal{L}_{\Pi_i}^{\text{kin}}+\mathcal{L}_{\Pi_i}^{\text{int}}}\, .
\end{equation}
Here $\pi_i$ is a light field, which in our case will correspond to the different fields of the SM, and $\Pi_i$ is a heavy field $m_{\Pi_i}\sim \Lambda_\text{NP}$ belonging to a NP extension, which can correspond to a scalar $\Phi_i$, fermion $\Psi_i$ or vector $V_i$. The superscript \textit{kin} denotes kinetic terms and \textit{int} interaction terms, where the last term in Eq.~\eqref{eq:lhLag} also includes the interactions between heavy and light fields. By assuming that we are only interested in observable phenomena below some heavy scale $\Lambda_\text{NP}$, we may rewrite the interaction Lagrangian involving heavy fields in terms of effective operators $\mathcal{L}_{\Pi_j}^{\text{int}} \to \mathcal{L}_\text{eff}$ where $\mathcal{L}_\text{eff}$ corresponds to the EFT Lagrangian given in Eq.~\eqref{eq:wilsonLNV}. This replacement can be done via covariant derivative expansion (CDE) (for details see Appendix~\ref{sec:CDE}), in which we take the derivative of a general interaction Lagrangian with respect to a heavy field~\cite{Henning:2014wua,Gargalionis:2020xvt}
\begin{equation}
\label{eq:derivative}
	\frac{\partial \mathcal{L}_{\Pi_i}^{\text{int}}}{\partial \Pi_i^\dagger}=\sum_{j,a,b,c}\lr{1+ c^{\Pi_j}\frac{\partial \mathcal{L}_{\Pi_j}^{\text{int}}}{\partial \Pi^\dagger_i}}\lr{c^\pi_2\pi_{a}\pi_{b}+c^{\pi}_3\pi_{a}\pi_b\pi_c}\, ,
\end{equation}
where $c^{\Pi_i}$ has mass dimension $d=3$, and $c_2^\pi$, $c_3^\pi$ have the appropriate mass dimensions depending on whether the light fields $\pi_i$ are fermions or bosons\footnote{Note that the term with $c_3^\pi$ is only meaningful if all of the fields $\pi_{a}$, $\pi_b$, and $\pi_c$ are bosons.}. To end up with a Lagrangian with only light fields, Eq.~\eqref{eq:derivative} can then be applied again onto itself by expanding the term with the remaining derivative ${\partial \mathcal{L}_{\Pi_j}^{\text{int}}}/{\partial \Pi^\dagger_j}$ in the same way, repeating this step until only light fields remain. This fully light Lagrangian can then be implemented in the expansions shown below in order to obtain a series of effective operators. For a scalar field, we have the effective Lagrangian~\cite{Henning:2014wua,Gargalionis:2020xvt}
\begin{equation}
	\label{eq:LeffScalar}
	\mathcal{L}_{\text{eff}}^\Phi = \frac{\partial \mathcal{L}_\Phi^{\text{int}}}{\partial \Phi}\lr{\frac{1}{m_\Phi^2}-\frac{D^2}{m_\Phi^4} + \dots}\frac{\partial \mathcal{L}_\Phi^{\text{int}}}{\partial \Phi^*}\, .
\end{equation}
For Dirac fermions we have~\cite{Henning:2014wua,Gargalionis:2020xvt}
\begin{equation}
\label{eq:LeffFermion}
	\begin{aligned}
		\mathcal{L}_{\text{eff}}^\Psi &=\frac{\partial \mathcal{L}_\Psi^{\text{int}}}{\partial \bar{\Psi}}\lr{\frac{1}{m_\Psi^2}+\frac{D^2+\tfrac{1}{2}X_{\mu\nu}\sigma^{\mu\nu}}{m_\Psi^4}+\dots}\left(i\slashed D+m_\Psi\right)\frac{\partial \mathcal{L}_\Psi^{\text{int}}}{\partial \Psi}\, ,
	\end{aligned}
\end{equation}
where $\sigma^{\mu\nu}\equiv\frac{i}{2}\left[\gamma^\mu,\gamma^\nu\right]$ corresponds to a tensor current, and where $X_{\mu\nu}$ is the field strength of a gauge field corresponding to the covariant derivative $D$.
The relation for Majorana fermions can be obtained via the replacement $m_\Psi\,\to\, \tfrac{1}{2}m_\Psi$. Lastly, for vector fields, we have
\begin{equation}
	\begin{aligned}
		\mathcal{L}_{\text{eff}}^{V} &=\frac{\partial \mathcal{L}_{V}^{\text{int}}}{\partial V^{\mu}}\lr{\frac{\eta^{\mu\nu}}{m_{V}^2}+\frac{D^2\eta^{\mu\nu} -D^\mu D^{\nu}}{m_{V}^4}+\dots}\frac{\partial \mathcal{L}_{V}^{\text{int}}}{\partial V^{\nu *}}\, .
	\end{aligned}
\end{equation}

For each dimension-7 $\Delta L = 2$ SMEFT operator, we systematically combine the SM fields in pairs and read off the corresponding representation under the SM gauge group for each combination. This then corresponds to the representation of an NP field that would couple to this pair in a vertex belonging to an NP interaction Lagrangian. In turn, this NP field can then be paired with the remaining SM fields until only renormalisable terms are left, at which point we identify the sum of these terms with the interaction Lagrangian that is our starting point in the CDE. By systematically repeating this procedure, we obtain a list of all possible tree-level UV completions. This lets us match each operator to a finite set of tree-level UV completions, including operators with derivatives. We limit ourselves to tree-level UV completions in this work.

\begin{table}[]
\centering
\begin{tabular}{c|l|l}
	Field & Rep & Coupling to SM fields ($+\hc$)\\[0mm]
	\specialrule{1pt}{0pt}{0pt}
	\raisebox{-1pt}{$\mathcal{S}$}  & \raisebox{-1pt}{$S(1,1,0)$}  & \raisebox{-1pt}{$\frac{1}{2}\kappa_{\mathcal{S}}\mathcal{S}H^\dagger H + \frac{1}{2}\lambda_{\mathcal{S}}\mathcal{S}\mathcal{S}H^\dagger H$} \\[0mm]
	$\Xi$  & $S(1,3,0)$  &  $\frac{1}{2}\kappa_\Xi H^\dagger\Xi^a\sigma^a H + \frac{1}{2}\lambda_\Xi\lr{\Xi\Xi}\lr{H^\dagger H}$ \\[0mm]
	$h$  & $S(1,1,1)$  & $y_h h^\dagger \bar{L}i\sigma_2 L^c + \kappa_h h^\dagger \tilde{H}^\dagger H$ \\[0mm]
	$\Delta$  & $S(1,3,1)$  & $\frac{1}{4}\lambda_{\Delta}\lr{\Delta^\dagger\Delta}\lr{H^\dagger H}+\frac{1}{4}\lambda'_{\Delta}f_{abc}\lr{\Delta^{a\dagger}\Delta^b}\lr{H^\dagger\sigma^c H}$\\[0mm]
	& & $+ y_{\Delta}\Delta^{a\dagger}\bar{L}\sigma^a i\sigma_2 L^c+\kappa_\Delta\Delta^{a\dagger}\big(\tilde{H}^\dagger\sigma^a H\big)$\\[0mm]
	$\varphi$  & $S(1,2,1/2)$  & $\lambda_\varphi\lr{\varphi^\dagger H}\lr{H^\dagger H}+y_\varphi^e\varphi^\dagger\bar{e}L+y_\varphi^d\varphi^\dagger\bar{d}Q+y_\varphi^u\varphi^\dagger i\sigma_2\bar{Q}^T u $\\[0mm]
	$\Theta_1$  & $S(1,4,1/2)$  & $\lambda_{\Theta_1}\lr{H^\dagger \sigma^a H}C^I_{ab}\tilde{H}^b\epsilon_{IJ}\Theta_1^J$ \\[0mm]
	$\Theta_3$  & $S(1,4,3/2)$  & $\lambda_{\Theta_3}\big(H^\dagger \sigma^a \tilde{H}\big)C^I_{ab}\tilde{H}^b\epsilon_{IJ}\Theta_3^J$ \\[0mm]
	$S_1$  & $S(\bar{3},1,1/3)$  & $y_{S_1}^{ql} S_1 \bar{Q}^c i\sigma_2 L+y_{S_1}^{qq} S_1\bar{Q}i\sigma_2 Q^c +y_{S_1}^{du} S_1\bar{d}u^c+y_{S_1}^{eu} S_1\bar{e}^cu$ \\[0mm]
	$\tilde{R}_2$  & $S(3,2,1/6)$  & $y_{\tilde{R}_2}\tilde{R}_2^\dagger i\sigma_2\bar{L}^T d $ \\[0mm]
	$S_3$  & $S(\bar{3},3,1/3)$  & $y_{S_3}^{ql}S_3^a \bar{Q}^c i\sigma_2\sigma^a L+y_{S_3}^{qq}S_3^a \bar{Q}\sigma^a i\sigma_2 Q^c$ \\[0mm]
	$N$  & $F(1,1,0)$  & $\lambda_N\bar{N}\tilde{H}^\dagger L$ \\[0mm]
	$\Sigma$  & $F(1,3,0)$  & $\frac{1}{2}\lambda_\Sigma\bar{\Sigma}^a\tilde{H}^\dagger \sigma^a L$ \\[0mm]
	$\Sigma_1$  & $F(1,3,-1)$  & $\frac{1}{2}\lambda_{\Sigma_1}\bar{\Sigma}_1^aH^\dagger \sigma^a L$ \\[0mm]
	$\Delta_1$  & $F(1,2,-1/2)$  & $\lambda_{\Delta_1}\bar{\Delta}_1 He$ \\[0mm]
	$\Delta_3$  & $F(1,2,-3/2)$  & $\lambda_{\Delta_3}\bar{\Delta}_3 \tilde{H}e$ \\[0mm]
	$F_4$  & $F(1,4,1/2)$  & $\--$ \\[0mm]
	$U$  & $F(3,1,2/3)$  & $\lambda_U\bar{U}\tilde{H}^\dagger Q$ \\[0mm]
	$Q_5$  & $F(3,2,-5/6)$  & $\lambda_{Q_5}\bar{Q}_5\tilde{H}d$ \\[0mm]
	$Q_7$  & $F(3,2,7/6)$  & $\lambda_{Q_7} \bar{Q}_7 H u$ \\[0mm]
	$T_1$  & $F(3,3,-1/3)$  & $\frac{1}{2}\lambda_{T_1}\bar{T}_1^aH^{\dagger}\sigma^a Q^b$ \\[0mm]
	$T_2$  & $F(3,3,2/3)$  &  $\frac{1}{2}\lambda_{T_2}\bar{T}_2^a\tilde{H}^{\dagger}\sigma^a Q^b$ \\[0mm]
	$W'_1$  & $V(1,1,1)$  &  $\frac{1}{2}g_{W'_1}^{du} {W'_1}^{\mu\dagger}\bar{d}\gamma_\mu u + g_{W'_1}^H {W'_1}^{\mu\dagger} iD_\mu H^T i\sigma_2 H$ \\[0mm]
	$V_{3}$  & $V(1,2,3/2)$  & $V_3^\mu \bar{e}^c\gamma_\mu L$ \\[0mm]
	$U_1$  & $V(3,1,2/3)$  & $g_{U_1}^{ed}U_1^{\mu\dagger}\bar{e}\gamma_\mu d+g_{U_1}^{lq}U_1^{\mu\dagger}\bar{L}\gamma_\mu Q$ \\[0mm]
	$\bar{V}_2$  & $V(\bar{3},2,-1/6)$  &  $g_{\bar{V}_2}^{ul}\bar{V}_2^\mu \bar{u}^c \gamma_\mu L+g_{\bar{V}_2}^{dq}\bar{V}_2^\mu\bar{d}\gamma_\mu i\sigma_2 Q^c$ \\[0mm]
	$U_3$  & $V(3,3,2/3)$  & $g_{U_3} U_3^{a\mu\dagger}\bar{L}\gamma_\mu \sigma^a Q$ \\[0mm]
\end{tabular}
\caption{A full list of the heavy BSM fields that appear in UV-completions of dimension-7 $\Delta L = 2$ operators. The second column depicts the representation under the SM gauge group, where $S$, $F$, and $V$ denote a scalar, fermion, and vector field, respectively. The third column shows the couplings to SM fields.}
\label{tab:D7SMcouplings}
\end{table}

\begin{table}[]
	\setlength\extrarowheight{5pt}
	\tabcolsep=0.11cm
	\centering
	\begin{tabular}{>{\centering}m{15pt}?>{\centering}m{50pt}?>{\centering}m{50pt}?>{\centering}m{50pt}?>{\centering}m{55pt}?>{\centering}m{50pt}?>{\centering}m{50pt}?>{\centering}m{50pt}?c}
		\multicolumn{1}{c}{\cellcolor{newgray} } &\multicolumn{1}{c}{\cellcolor{newgray}\raisebox{2pt}{ $\mathcal{O}_{LH}$ }}& \multicolumn{1}{c}{\cellcolor{newgray} \raisebox{2pt}{$\mathcal{O}_{LeHD}$} } & \multicolumn{1}{c}{\cellcolor{newgray} \raisebox{2pt}{$\mathcal{O}_{\bar{e}LLLH}$} } &  \multicolumn{1}{c}{\cellcolor{newgray}\raisebox{2pt}{$\mathcal{O}_{\bar{d}LueH}$ } } & \multicolumn{1}{c}{ \cellcolor{newgray}\raisebox{2pt}{ $\mathcal{O}_{\bar{d}LQLH1}$} } & \multicolumn{1}{c}{\cellcolor{newgray} \raisebox{2pt}{$\mathcal{O}_{\bar{d}LQLH2}$ }} & \multicolumn{1}{c}{\hspace{0pt}\cellcolor{newgray} \raisebox{2pt}{$\mathcal{O}_{\bar{Q}uLLH}$}\hspace{0pt} } &  \\
		\addlinespace[-1pt]\cmidrule[1.1pt](l{0pt}r{0pt}){2-8}\addlinespace[-4pt]
		\cellcolor{newgray} $\mathcal{S}$  & $\Delta,\!$ $N,\!$ $\Sigma\!$ &  &  &  &  &  &  &  \\
		\addlinespace[-1pt]\cmidrule[1.1pt](l{0pt}r{0pt}){2-8}\addlinespace[-4pt]
		\cellcolor{newgray} $\Xi$  & $\Delta,\!$ $\Sigma\!$ &  &  &  &  &  & &   \\
		\addlinespace[-1pt]\cmidrule[1.1pt](l{0pt}r{0pt}){2-8}\addlinespace[-4pt]
		\cellcolor{newgray} $h$  &  &  & $\varphi,\!$ $N,\!$ $\Delta_3^\dagger\!$ &  &  & $\varphi,\!$ $U,\!$ $Q_5^\dagger\!$ & &   \\
		\addlinespace[-1pt]\cmidrule[1.1pt](l{0pt}r{0pt}){2-8}\addlinespace[-4pt]
		\cellcolor{newgray} $\Delta$  & $\mathcal{S},\!$ $\Xi,\!$ $\Delta,\!$ $\varphi,\!$ $\Theta_1,\!$ $\Theta_3,\!$ $\Sigma\!$ & $\Sigma,\!$ $\Delta_1^\dagger\!$ & $\varphi,\!$ $\Sigma,\!$ $\Delta_3^\dagger\!$ &  & $\varphi,\!$ $Q_5^\dagger,\!$ $T_2\!$ &  & $\varphi,\!$ $Q_7,\!$ $T_1^\dagger\!$ &  \\
		\addlinespace[-1pt]\cmidrule[1.1pt](l{0pt}r{0pt}){2-8}\addlinespace[-4pt]
		\cellcolor{newgray} $\varphi$  & $\Delta,\!$ $N,\!$ $\Sigma\!$ &  & $h,\!$ $\Delta,\!$ $N,\!$ $\Sigma\!$ &  & $\Delta,\!$ $N,\!$ $\Sigma\!$ & $h,\!$ $\Sigma\!$ & $\Delta,\!$ $N,\!$ $\Sigma\!$ &  \\
		\addlinespace[-1pt]\cmidrule[1.1pt](l{0pt}r{0pt}){2-8}\addlinespace[-4pt]
		\cellcolor{newgray} $\Theta_1$  & $\Delta,\!$ $\Sigma\!$ &  &  &  &  &  &  &  \\
		\addlinespace[-1pt]\cmidrule[1.1pt](l{0pt}r{0pt}){2-8}\addlinespace[-4pt]
		\cellcolor{newgray} $\Theta_3$  & $\Delta,\!$ $\Sigma_1^\dagger\!$ &  &  &  &  &  &  &  \\
		\addlinespace[-1pt]\cmidrule[1.1pt](l{0pt}r{0pt}){2-8}\addlinespace[-4pt]
		\cellcolor{newgray} $S_1$  &  &  &  & $\tilde{R}_2,\!$ $N,\!$ $Q_5^\dagger\!$ & $\tilde{R}_2,\!$ $N,\!$ $Q_5^\dagger\!$ &  &  &  \\
		\addlinespace[-1pt]\cmidrule[1.1pt](l{0pt}r{0pt}){2-8}\addlinespace[-4pt]
		\cellcolor{newgray} $\tilde{R}_2$  &  &  &  & $S_1,\!$ $\Delta_1^\dagger,\!$ $Q_7\!$ & $S_1,\!$ $S_3,\!$ $N,\!$ $\Sigma,\!$ $T_2\!$ & $S_3,\!$ $\Sigma,\!$ $U\!$ & &   \\
		\addlinespace[-1pt]\cmidrule[1.1pt](l{0pt}r{0pt}){2-8}\addlinespace[-4pt]
		\cellcolor{newgray} $S_3$  &  &  &  &  & $\tilde{R}_2,\!$ $\Sigma,\!$ $Q_5^\dagger\!$ & $\tilde{R}_2\!$, $\Sigma,\!$ $Q_5^\dagger\!$ & &   \\
		\addlinespace[-1pt]\cmidrule[1.1pt](l{0pt}r{0pt}){2-8}\addlinespace[-4pt]
		\cellcolor{newgray} $N$  & $\mathcal{S},\!$ $\varphi,\!$ $\Delta_1^\dagger\!$ & $\Delta_1^\dagger\!$ & $h,\!$ $\varphi\!$ & $S_1,\!$ $W_1',\!$ $U_1\!$ & $\varphi,\!$ $S_1,\!$ $\tilde{R}_2\!$ &  & $\varphi,\!$ $U_1,\!$ $\bar{V}_2^\dagger\!$ &  \\
		\addlinespace[-1pt]\cmidrule[1.1pt](l{0pt}r{0pt}){2-8}\addlinespace[-4pt]
		\cellcolor{newgray} $\Sigma$  & $\mathcal{S},\!$ $\Xi,\!$ $\Delta,\!$ $\varphi,\!$ $\Theta_1,\!$ $\Delta_1^\dagger,\!$ $F_4\!$ & $\Delta,\!$ $\Delta_1^\dagger\!$ & $\Delta,\!$ $\varphi\!$ &  & $\varphi,\!$ $\tilde{R}_2,\!$ $S_3\!$ & $\varphi,\!$ $\tilde{R}_2,\!$ $S_3\!$ & $\varphi,\!$ $\bar{V}_2^\dagger,\!$ $U_3\!$ &  \\
		\addlinespace[-1pt]\cmidrule[1.1pt](l{0pt}r{0pt}){2-8}\addlinespace[-4pt]
		\cellcolor{newgray} $\Sigma_1^\dagger$  & $\Theta_3\!$ &  &  &  &  &  & &   \\
		\addlinespace[-1pt]\cmidrule[1.1pt](l{0pt}r{0pt}){2-8}\addlinespace[-4pt]
		\cellcolor{newgray} $\Delta_1^\dagger$  & $N,\!$ $\Sigma\!$ & $\Delta,\!$ $N,\!$ $\Sigma\!$ &  & $\tilde{R}_2,\!$ $W_1',\!$ $\bar{V}_2^\dagger\!$ &  &  & &   \\
		\addlinespace[-1pt]\cmidrule[1.1pt](l{0pt}r{0pt}){2-8}\addlinespace[-4pt]
		\cellcolor{newgray} $\Delta_3^\dagger$  &  &  & $h,\!$ $\Delta\!$ &  &  &  & &   \\
		\addlinespace[-1pt]\cmidrule[1.1pt](l{0pt}r{0pt}){2-8}\addlinespace[-4pt]
		\cellcolor{newgray} $F_4$  & $\Sigma\!$ &  &  &  &  &  & &   \\
		\addlinespace[-1pt]\cmidrule[1.1pt](l{0pt}r{0pt}){2-8}\addlinespace[-4pt]
		\cellcolor{newgray} $U$  &  &  &  &  &  & $h,\!$ $\tilde{R}_2\!$ & &   \\
		\addlinespace[-1pt]\cmidrule[1.1pt](l{0pt}r{0pt}){2-8}\addlinespace[-4pt]
		\cellcolor{newgray} $Q_5^\dagger$  &  &  &  & $S_1,\!$ $V_3,\!$ $\bar{V}^\dagger_2\!$ & $\Delta,\!$ $S_1,\!$ $S_3\!$ & $h,\!$ $S_3\!$ & &   \\
		\addlinespace[-1pt]\cmidrule[1.1pt](l{0pt}r{0pt}){2-8}\addlinespace[-4pt]
		\cellcolor{newgray} $Q_7$  &  &  &  & $\tilde{R}_2,\!$ $V_3,\!$ $U_1\!$ &  &  & $\Delta,\!$ $U_1,\!$ $U_3\!$ &   \\
		\addlinespace[-1pt]\cmidrule[1.1pt](l{0pt}r{0pt}){2-8}\addlinespace[-4pt]
		\cellcolor{newgray} $T_1^\dagger$  &  &  &  &  &  &  & $\Delta,\!$ $\bar{V}_2^\dagger\!$ &  \\
		\addlinespace[-1pt]\cmidrule[1.1pt](l{0pt}r{0pt}){2-8}\addlinespace[-4pt]
		\cellcolor{newgray} $T_2$  &  &  &  &  & $\Delta,\!$ $\tilde{R}_2\!$ &  & &   \\
		\addlinespace[-1pt]\cmidrule[1.1pt](l{0pt}r{0pt}){2-8}\addlinespace[-4pt]
		\cellcolor{newgray} $W_1'$  &  &  &  & $N,\!$ $\Delta_1^\dagger,\!$ $V_3\!$ &  &  & &   \\
		\addlinespace[-1pt]\cmidrule[1.1pt](l{0pt}r{0pt}){2-8}\addlinespace[-4pt]
		\cellcolor{newgray} $V_3$  &  &  &  & $Q_5^\dagger,\!$ $Q_7,\!$ $W_1'\!$ &  &  & &   \\
		\addlinespace[-1pt]\cmidrule[1.1pt](l{0pt}r{0pt}){2-8}\addlinespace[-4pt]
		\cellcolor{newgray} $U_1$  &  &  &  & $N,\!$ $Q_7,\!$ $\bar{V}_2^\dagger\!$ &  &  & $N,\!$ $Q_7,\!$ $\bar{V}_2^\dagger\!$ &  \\
		\addlinespace[-1pt]\cmidrule[1.1pt](l{0pt}r{0pt}){2-8}\addlinespace[-4pt]
		\cellcolor{newgray} $\bar{V}_2^\dagger$  &  &  &  & $\Delta_1^\dagger,\!$ $Q_5^\dagger,\!$ $U_1\!$ &  &  &  $N,\!$ $\Sigma,\!$ $T_1^\dagger,\!$ $U_1,\!$ $U_3\!$ &  \\
		\addlinespace[-1pt]\cmidrule[1.1pt](l{0pt}r{0pt}){2-8}\addlinespace[-4pt]
		\cellcolor{newgray} $U_3$  &  &  &  &  &  &  & $\Sigma,\!$ $Q_7,\!$ $\bar{V}_2^\dagger\!$ &  \\
		\addlinespace[-1pt]\cmidrule[1.1pt](l{0pt}r{0pt}){2-8}\addlinespace[-4pt]
	\end{tabular}
	\caption{Combinations of BSM fields (c.f.\ Tab.~\ref{tab:D7SMcouplings}) that realize the different dimension-7 $\Delta L = 2$ operators at tree-level. The fields to the left can be combined with either of the fields in the same row to realize the operator shown at the top of the corresponding column. Each field has its own row and appears also in the entries in other rows, such that each combination leading to a distinct simplified model is listed exactly twice.}
	\label{tab:bigtab23}
\end{table}

In Tab.~\ref{tab:D7SMcouplings} we list the different BSM fields that appear in the tree-level UV-completions of the dimension-7 $\Delta L=2$ operators. Here the representation of each field is shown with respect to the SM gauge group $SU(3)_c\times SU(2)_L\times U(1)_Y$, and $S$, $F$, and $V$ denote whether the field is a scalar, fermion, or vector, respectively. The rightmost column shows the different couplings to the SM fields, where $\tilde{H}=i\sigma_2 H^*$. Each UV-completion consists of two fields, such that a field in the left grey column should be combined with a single field from one of the cells in the corresponding row. Note that two or more BSM fields can also couple to each other in a single vertex. Note also that $F_4$ does not have any coupling to only SM fields. It appears in a UV-completion of $\mathcal{O}_{LH}$ together with $\Sigma$, which $F_4$ can couple to together with the Higgs. This UV-completion then involves two instances of $\Sigma$ such that the whole diagram resembles that of the inverse seesaw.

In Tab.~\ref{tab:bigtab23} we show the different combinations of BSM fields from Tab.~\ref{tab:D7SMcouplings} that lead to tree-level UV-completions of the dimension-7 $\Delta L = 2$ operators. Here, the field shown in the leftmost column can be combined with one of the other fields of the same row to realise the operator shown at the top of the corresponding column.

In this work, we only consider UV-completions for seven out of the original twelve dimension-7 $\Delta L = 2$ operators, namely those of types $\Psi^4H$, $\Psi^2H^4$, and $\Psi^2H^3D$. As can be seen from Tab.~\ref{tab:dim7op}, operators that contain $D^2$ or $X$ can be understood as higher-order corrections to a lower-dimensional operator, which in our case would correspond to the dimension-5 operator. Furthermore, as was noted in Ref.~\cite{Gargalionis:2020xvt}, since the $\Psi^4D$-type operator $\mathcal{O}_{\bar duLLD}$ contains a single covariant derivative, it must originate from a UV-completion containing a fermion mediator, c.f.\ Eq.~\eqref{eq:LeffFermion}. However, since it has four external fermion legs, there is no such UV-completion at tree-level, and we conclude that no such tree-level UV-completion is possible for this operator. 

In what follows, we still aim to focus on UV-completions that do not also lead to the dimension-5 operator at tree-level, since if this operator is induced, it would likely dominate over any dimension-7 contribution. Note that even when it comes to the considered operators, the Weinberg operator can be generated at tree-level if either of the type-I or type-III seesaw fields $N$ or $\Sigma$ are present in a given simplified model, or if the type-II seesaw field $\Delta$ is present with both couplings to a lepton doublet pair and a Higgs pair. While a UV-completion with either $N$ or $\Sigma$ does generate the Weinberg operator, it could be the case that their coupling to the SM is suppressed, and a higher dimensional operator would dominate. However, this higher dimensional operator would be of dimension-9 and not -7. The reason for this is that any $\Delta L = 2$ dimension-7 operator containing at least one Higgs field and two lepton doublets\footnote{A similar argument can be made for the case where one of the lepton doublets is exchanged for a singlet.} can be written as $LH\times L\mathcal{O}^{(3)}$, where $\mathcal{O}^{(3)}$ denotes a three-dimensional object. The factor $LH$ can couple to one of the seesaw fermions $F\in\{N,\Sigma\}$, such that the two parts $LH$ and $L\mathcal{O}^{(3)}$ of the dimension-7 operator are connected via this field. If the $FLH$ coupling is large, it can be used twice to give a dimension-5 operator of the form $LHLH$, which is exactly the Weinberg operator. We then expect this operator to dominate over the dimension-7 operator. However, if the $FL\mathcal{O}^{(3)}$ coupling is significantly larger, we can instead use this coupling twice to get a dimension-9 operator of the form $L\mathcal{O}^{(3)}L\mathcal{O}^{(3)}$, which should dominate over the dimension-7 operator since it does not contain the suppressed $FLH$ coupling. In neither scenario do we have a dimension-7 operator providing the dominant contribution. As dimension-9 operators are beyond the scope of our analysis, we do not further evaluate the UV-completions containing $N$ or $\Sigma$ in the following sections, except simply to note down how and in which operators they appear. Note that since the type-II seesaw field $\Delta$ couples twice to the SM, via $\Delta HH$ and $\Delta LL$, it does not generate the dimension-5 operator if only one coupling constant is non-vanishing, and it can therefore participate in dominant dimension-7 operators. However, the corresponding neutrino mass diagrams are often topologically equivalent to radiative corrections of the type-II seesaw diagram.

%%%%%%%%%%%%%%%%%%%%%%%%%%%%%%%%%%%%%%%%%%%%%%%%%%%%%%%%%%%%%%%%%%%%%%%%%%%%%%%%%
%%%%%%%%%%%%%%%%%%%%%%%%%%%%%%%%%%%%%%%%%%%%%%%%%%%%%%%%%%%%%%%%%%%%%%%%%%%%%%%%%

\section{Neutrino masses without tree-level dim-5 Weinberg operator}
\label{sec:EFTvsUVmass}
%%%%%%%%%%%%%%%%%%%%%%%%%%%%%%%%%%%%%%%%%%%%%%%%%%%%%%%%%%%%%%%%%%%%%%%%%%%%%%%%%

In this section, we discuss the different neutrino mass topologies that can be generated in the tree-level UV completions of the dimension-7 $\Delta L = 2$ operators. The operators that generate such tree-level UV-completions can be divided into three classes: $\Psi^2 H^4$, $\Psi^2 H^3 D$, and $\Psi^4 H$, where $\Psi$ denotes a fermion, $H$ the Higgs field, and $D$ a covariant derivative.

\subsection*{$\Psi^4 H$}\label{sec:EFTvsUVmass4f}
\begin{figure}
	\centering
	\includegraphics[width=0.45\textwidth]{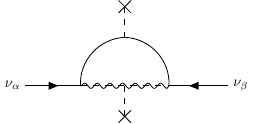}
	\includegraphics[width=0.45\textwidth]{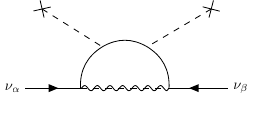}
	\caption{Radiative neutrino mass realised via topology I (left) and topology II (right).}
	\label{fig:topsIandII}
\end{figure}

Operators of type $\Psi^4 H$ lead to two topologies for 1-loop radiative neutrino masses, as shown in Fig.~\ref{fig:topsIandII}. In topology I (Fig.~\ref{fig:topsIandII} left), two new bosonic fields are introduced, which subsequently mix via a dimensionful coupling to the Higgs field. In topology II (Fig.~\ref{fig:topsIandII} right), there is one new boson and one (vector-like) fermion. For $N$ or $\Sigma$ mediators, there is a third topology with a Higgs vacuum expectation value (vev) insertion on a lepton leg outside the loop. Neutrino mass diagrams generated from UV-completions of operators of this type will contain at least one loop, since we need to close off two out of the four external fermion legs in order to get a self-energy type diagram for the remaining two. The bosonic mediators may be scalar or vector, depending on whether they contract a spinor with an antispinor (vector), or if they contract two spinors or antispinors (scalar). If the bosonic mediator in topology II is the type-II seesaw field $\Delta$ there is an additional diagram for operator $\mathcal{O}_{\bar eLLLH}$ where the fermion loop is disconnected from the neutrinos and the two fermion lines are coupled via $\Delta$. This can be seen as a vertex correction to the regular type-II seesaw diagram. 

\subsection*{$\Psi^2H^3 D$}\label{sec:EFTvsUVmass2f3HD}
There is only a single dimension-7 $\Delta L = 2$ operator of type $\Psi^2H^3 D$, namely $\mathcal{O}_{LeHD}$. Furthermore, this operator only has a single tree-level UV-completion, which does not lead to the Weinberg operator, namely the combination of $\Delta$ and $\Delta_1^\dagger$ leading to topology I. Unlike the $\Psi^4 H$-type operators, $\mathcal{O}_{LeHD}$ contains only one lepton doublet. In order to generate a neutrino mass, the lepton singlet must be converted into a doublet via a Higgs interaction, and this is done by closing a loop with one of the three external Higgs fields, such that there are only two insertions of the Higgs vev, and topology I is generated. 

\subsection*{$\Psi^2H^4$}\label{sec:EFTvsUVmass2f4H}
\begin{figure}
	\centering
	\includegraphics[width=0.40\textwidth]{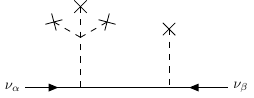}
	\includegraphics[width=0.25\textwidth]{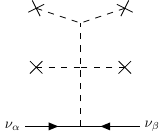}
	\includegraphics[width=0.25\textwidth]{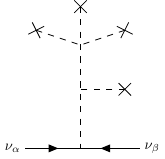}
	\caption{Tree-level neutrino mass realised via topology III (left), topology IV (centre) and topology V (right).}
	\label{fig:topIII}
\end{figure}

For type $\Psi^2H^4$ there is again only a single operator, namely $\mathcal{O}_{LH}$~Refs.~\cite{Cepedello:2017lyo,Cepedello:2017eqf}. Such a simple extension leads to a wide range of tree-level UV completions containing two or three new fields. For consistency with the other cases, we will focus only on UV-completions with two fields.
\vspace{-2.0 cm}
\begin{table}[H] \centering
\begin{tabular}{c c}
%%%%%%%%%%%%%%%%%%%
$\mathcal{O}_{LH}=\epsilon_{ij}\epsilon_{mn}\lrs{\overline{L_p^c}{}^iL_r^m}H^jH^n\lrs{H^\dagger H}$ 
&
$\mathcal{O}_{LeHD}=\epsilon_{ij}\epsilon_{mn}\lrs{\overline{L_p^c}{}^i\gamma_\mu e_r}H^j\lrs{H^miD^\mu H^n}$\\
    \begin{minipage}[t]{0.50\textwidth}\centering

    \resizebox{1.\textwidth}{!}{
\begin{tabular}{>{\centering}m{10pt} | >{\centering}m{10pt} | >{\centering}m{10pt} | >{\centering}m{10pt} | >{\centering}m{10pt} | >{\centering}m{10pt} | >{\centering}m{10pt} | >{\centering}m{10pt} | >{\centering}m{10pt} | >{\centering}m{10pt} | >{\centering}m{10pt} | c |}
	\cellcolor{newgray} & \cellcolor{newgray} $\mathcal{S}$ & \cellcolor{newgray} $\Xi$ & \cellcolor{newgray} $\Delta$ & \cellcolor{newgray} $\varphi$ & \cellcolor{newgray} $\Theta_1$ & \cellcolor{newgray} $\Theta_3$ & \cellcolor{newgray} $N$ & \cellcolor{newgray} $\Sigma$ & \cellcolor{newgray} $\Sigma_1^\dagger$ & \cellcolor{newgray} $\Delta_1^\dagger$ & \cellcolor{newgray} $F_4$ \\
	\specialrule{0.8pt}{0pt}{0pt}
 \cellcolor{newgray} $\mathcal{S}$ &  &  & {\small IV} &  &  &  & {\small $\Circle$} & {\small $\Circle$} &  &  &  \\
	\specialrule{0.8pt}{0pt}{0pt}
 \cellcolor{newgray} $\Xi$ &  &  & {\small IV} &  &  &  &  & {\small $\Circle$} &  &  &  \\
	\specialrule{0.8pt}{0pt}{0pt}
 \cellcolor{newgray} $\Delta$ & {\small IV} & {\small IV} & {\small $\Circle$} & \mbox{\small V} & \mbox{\hspace{-1mm} \small V} & \mbox{\small V} &  & {\small $\Circle$} &  &  &  \\
	\specialrule{0.8pt}{0pt}{0pt}
 \cellcolor{newgray} $\varphi$ &  &  & \mbox{\small V} &  &  &  & {\small $\Circle$} & {\small $\Circle$} &  &  &  \\
	\specialrule{0.8pt}{0pt}{0pt}
 \cellcolor{newgray} $\Theta_1$ &  &  & \mbox{\hspace{-1mm} \small V} &  &  &  &  & {\small $\Circle$} &  &  &  \\
	\specialrule{0.8pt}{0pt}{0pt}
 \cellcolor{newgray} $\Theta_3$ &  &  & \mbox{\small V} &  &  &  &  &  & {\small III} &  &  \\
	\specialrule{0.8pt}{0pt}{0pt}
 \cellcolor{newgray} $N$ & {\small $\Circle$} &  &  & {\small $\Circle$} &  &  &  &  &  & {\small $\Circle$} &  \\
	\specialrule{0.8pt}{0pt}{0pt}
 \cellcolor{newgray} $\Sigma$ & {\small $\Circle$} & {\small $\Circle$} & {\small $\Circle$} & {\small $\Circle$} & {\small $\Circle$} &  &  &  &  & {\small $\Circle$} & {\small $\Circle$} \\
	\specialrule{0.8pt}{0pt}{0pt}
 \cellcolor{newgray} $\Sigma_1^\dagger$ &  &  &  &  &  & {\small III} &  &  &  &  &  \\
	\specialrule{0.8pt}{0pt}{0pt}
 \cellcolor{newgray} $\Delta_1^\dagger$ &  &  &  &  &  &  & {\small $\Circle$} & {\small $\Circle$} &  &  &  \\
	\specialrule{0.8pt}{0pt}{0pt}
 \cellcolor{newgray} $F_4$ &  &  &  &  &  &  &  & {\small $\Circle$} &  &  &  \\
	\specialrule{0.8pt}{0pt}{0pt}
\end{tabular}
    }
    \label{tab:LH}
    \end{minipage}%\\ \vspace{0.5 cm}
&
    \begin{minipage}[b]{0.50\textwidth}\centering
\begin{tabular}{c}
 \begin{minipage}[t]{0.40\textwidth}\centering
    \resizebox{1.\textwidth}{!}{
	\begin{tabular}{>{\centering}m{10pt} | >{\centering}m{10pt} |>{\centering}m{10pt} |>{\centering}m{10pt} | c |}
		\cellcolor{newgray} & \cellcolor{newgray} $\Delta$ & \cellcolor{newgray} $N$ & \cellcolor{newgray} $\Sigma$ & \hspace{-1pt}\cellcolor{newgray} $\Delta_1^\dagger$\hspace{-1pt}  \\
		\specialrule{0.8pt}{0pt}{0pt}
		\cellcolor{newgray} $\Delta$  &  &  & ${\small \Circle}$ & {\small \CIRCLE} \\
		\specialrule{0.8pt}{0pt}{0pt}
		\cellcolor{newgray} $N$   &  &  &  & ${\small \Circle}$ \\
		\specialrule{0.8pt}{0pt}{0pt}
		\cellcolor{newgray} $\Sigma$    & ${\small \Circle}$ &  &  & ${\small \Circle}$ \\
		\specialrule{0.8pt}{0pt}{0pt}
		\cellcolor{newgray} $\Delta_1^\dagger$    & {\small \CIRCLE} & ${\small \Circle}$ & ${\small \Circle}$ &  \\
		\specialrule{0.8pt}{0pt}{0pt}
	\end{tabular}
    }
    	\label{tab:LeHD}
    \end{minipage}
     \\
   $\mathcal{O}_{\bar eLLLH}=\epsilon_{ij}\epsilon_{mn}\lrs{\overline{e_p}L_r^i}\lrs{\overline{L_s^c}{}^j L_t^m}H^n$
     \\  
    \begin{minipage}[b]{0.60\textwidth}\centering
     \resizebox{1.\textwidth}{!}{
	\begin{tabular}{>{\centering}m{10pt}|>{\centering}m{10pt}|>{\centering}m{10pt}|>{\centering}m{10pt}|>{\centering}m{10pt}|>{\centering}m{15pt}|c|}
		\cellcolor{newgray} & \cellcolor{newgray} $h$ & \cellcolor{newgray} $\Delta$ & \cellcolor{newgray} $\varphi$ & \cellcolor{newgray} $N$ & \cellcolor{newgray} $\Sigma$ & \hspace{-1pt}\cellcolor{newgray} $\Delta_3^\dagger$\hspace{-1pt} \\
		\hline
		\cellcolor{newgray} $h$ &  &  & {\small I} & ${\small \Circle}$ &  & {\small II} \\
		\hline
		\cellcolor{newgray} $\Delta$ &  &  & {\small \CIRCLE} &  & ${\small \Circle}$ & {\small II} \\
		\hline
		\cellcolor{newgray} $\varphi$ & {\small I} & {\small \CIRCLE} &  & ${\small \Circle}$ & ${\small \Circle}$ &  \\
		\hline
		\cellcolor{newgray} $N$ & ${\small \Circle}$ &  & ${\small \Circle}$ &  &  &  \\
		\hline
		\cellcolor{newgray} $\Sigma$ &  & ${\small \Circle}$ & ${\small \Circle}$ &  &  &  \\
		\hline
		\cellcolor{newgray} $\Delta_3^\dagger$ & {\small II} & {\small II} &  &  &  &  \\
		\hline
\end{tabular}

    }
    \end{minipage}
    \end{tabular}
    \end{minipage}

     \vspace{0.5 cm}\\ 
    $\mathcal{O}_{\bar dLueH}=\epsilon_{ij}\lrs{\overline{d_p}L_r^i}\lrs{\overline{u_s^c}e_t}H^j$ 
    &
    $\mathcal{O}_{\bar dLQLH1}=\epsilon_{ij}\epsilon_{mn}\lrs{\overline{d_p}L_r^i}\lrs{\overline{Q_s^c}{}^jL_t^m}H^n$\\

    \begin{minipage}[b]{0.50\textwidth}\centering
    \resizebox{1.\textwidth}{!}{
	\begin{tabular}{>{\centering}m{10pt}|>{\centering}m{10pt}|>{\centering}m{10pt}|>{\centering}m{10pt}|>{\centering}m{10pt}|>{\centering}m{10pt}|>{\centering}m{10pt}|>{\centering}m{10pt}|>{\centering}m{10pt}|>{\centering}m{10pt}|c|}
		\cellcolor{newgray} & \cellcolor{newgray} $S_1$ & \cellcolor{newgray} $\tilde{R}_2$ & \cellcolor{newgray} $N$ & \cellcolor{newgray} $\Delta_1^\dagger$ & \cellcolor{newgray} $Q_5^\dagger$ & \cellcolor{newgray} $Q_7$ & \cellcolor{newgray} $W'_1$ & \cellcolor{newgray} $V_3$ & \cellcolor{newgray} $U_1$ & \cellcolor{newgray} $\bar{V}_2^\dagger$ \\
		\hline
		\cellcolor{newgray} $S_1$ &  & {\small I*} & ${\small \Circle}$ &  & {\small II*} &  &  &  &  &  \\
		\hline
		\cellcolor{newgray} $\tilde{R}_2$ & {\small I*} &  &  & {\small II*} &  & {\small II*} &  &  &  &  \\
		\hline
		\cellcolor{newgray} $N$ & ${\small \Circle}$ &  &  &  &  &  & ${\small \Circle}$ &  & ${\small \Circle}$ &  \\
		\hline
		\cellcolor{newgray} $\Delta_1^\dagger$ &  & {\small II*} &  &  &  &  & {\small II*} &  &  & {\small II*} \\
		\hline
		\cellcolor{newgray} $Q_5^\dagger$ & {\small II*} &  &  &  &  &  &  & {\small II*} &  & {\small II*} \\
		\hline
		\cellcolor{newgray} $Q_7$ &  & {\small II*} &  &  &  &  &  & {\small II*} & {\small II*} &  \\
		\hline
		\cellcolor{newgray} $W'_1$ &  &  & ${\small \Circle}$ & {\small II*} &  &  &  & {\small I*} &  &  \\
		\hline
		\cellcolor{newgray} $V_3$ &  &  &  &  & {\small II*} & {\small II*} & {\small I*} &  &  &  \\
		\hline
		\cellcolor{newgray} $U_1$ &  &  & ${\small \Circle}$ &  &  & {\small II*} &  &  &  & {\small I*} \\
		\hline
		\cellcolor{newgray} $\bar{V}_2^\dagger$ &  &  &  & {\small II*} & {\small II*} &  &  &  & {\small I*} &  \\
		\hline
\end{tabular}

    }
    \end{minipage}
  &
    \begin{minipage}[b]{0.50\textwidth}\centering
        \begin{minipage}[b]{0.90\textwidth}\centering
    \resizebox{1.\textwidth}{!}{
	\begin{tabular}{>{\centering}m{10pt}|>{\centering}m{10pt}|>{\centering}m{10pt}|>{\centering}m{10pt}|>{\centering}m{10pt}|>{\centering}m{10pt}|>{\centering}m{10pt}|>{\centering}m{10pt}|>{\centering}m{10pt}|c|}
		\cellcolor{newgray} & \cellcolor{newgray} $\Delta$ & \cellcolor{newgray} $\varphi$ & \cellcolor{newgray} $S_1$ & \cellcolor{newgray} $\tilde{R}_2$ & \cellcolor{newgray} $S_3$ & \cellcolor{newgray} $N$ & \cellcolor{newgray} $\Sigma$ & \cellcolor{newgray} $Q_5^\dagger$ & \cellcolor{newgray} $T_2$ \\
		\hline
		\cellcolor{newgray} $\Delta$ &  & {\small \CIRCLE} &  &  &  &  &  & {\small \CIRCLE} & {\small \CIRCLE} \\
		\hline
		\cellcolor{newgray} $\varphi$ & {\small \CIRCLE} &  &  &  &  & ${\small \Circle}$ & ${\small \Circle}$ &  &  \\
		\hline
		\cellcolor{newgray} $S_1$ &  &  &  & {\small I} &  & ${\small \Circle}$ &  & {\small II} &  \\
		\hline
		\cellcolor{newgray} $\tilde{R}_2$ &  &  & {\small I} &  & {\small I} & ${\small \Circle}$ & ${\small \Circle}$ &  & {\small II} \\
		\hline
		\cellcolor{newgray} $S_3$ &  &  &  & {\small I} &  &  & ${\small \Circle}$ & {\small II} &  \\
		\hline
		\cellcolor{newgray} $N$ &  & ${\small \Circle}$ & ${\small \Circle}$ & ${\small \Circle}$ &  &  &  &  &  \\
		\hline
		\cellcolor{newgray} $\Sigma$ &  & ${\small \Circle}$ &  & ${\small \Circle}$ & ${\small \Circle}$ &  &  &  &  \\
		\hline
		\cellcolor{newgray} $Q_5^\dagger$ & {\small \CIRCLE} &  & {\small II} &  & {\small II} &  &  &  &  \\
		\hline
		\cellcolor{newgray} $T_2$ & {\small \CIRCLE} &  &  & {\small II} &  &  &  &  &  \\
		\hline
\end{tabular}

    }
      \end{minipage}
     \end{minipage}

    \vspace{0.5 cm} \\ 
       $\mathcal{O}_{\bar dLQLH2}=\epsilon_{im}\epsilon_{jn}\lrs{\overline{d_p}L_r^i}\lrs{\overline{Q_s^c}{}^jL_t^m}H^n$ 
       &
       $\mathcal{O}_{\bar QuLLH}=\epsilon_{ij}\lrs{\overline{Q_p}u_r}\lrs{\overline{L_s^c}L_t^i}H^j$ \\
     
    \begin{minipage}[b]{0.50\textwidth}\centering
     \begin{minipage}[b]{0.70\textwidth}\centering
    \resizebox{1.\textwidth}{!}{
	\begin{tabular}{>{\centering}m{10pt}|>{\centering}m{10pt}|>{\centering}m{10pt}|>{\centering}m{10pt}|>{\centering}m{10pt}|>{\centering}m{10pt}|>{\centering}m{10pt}|c|}
   \cellcolor{newgray}  & \cellcolor{newgray} $h$ & \cellcolor{newgray} $\varphi$ & \cellcolor{newgray} $\tilde{R}_2$ & \cellcolor{newgray} $S_3$ & \cellcolor{newgray} $\Sigma$ & \cellcolor{newgray} $U$ & \cellcolor{newgray} $Q_5^\dagger$ \\
    \hline
   \cellcolor{newgray}  $h$ &  & {\small I*} &  &  &  & {\small II*} & {\small II*} \\
    \hline
    \cellcolor{newgray} $\varphi$ & {\small I*} &  &  &  & ${\small \Circle}$ &  &  \\
    \hline
    \cellcolor{newgray} $\tilde{R}_2$ &  &  &  & {\small I*} & ${\small \Circle}$ & {\small II*} &  \\
    \hline
    \cellcolor{newgray} $S_3$ &  &  & {\small I*} &  & ${\small \Circle}$ &  & {\small II*} \\
    \hline
    \cellcolor{newgray} $\Sigma$ &  & ${\small \Circle}$ & ${\small \Circle}$ & ${\small \Circle}$ &  &  &  \\
    \hline
    \cellcolor{newgray} $U$ & {\small II*} &  & {\small II*} &  &  &  &  \\
    \hline
    \cellcolor{newgray} $Q_5^\dagger$ & {\small II*} &  &  & {\small II*} &  &  &  \\
    \hline
\end{tabular}

    }
     \end{minipage}
         \end{minipage}

    &
             \begin{minipage}[b]{0.50\textwidth}\centering
              \begin{minipage}[b]{0.90\textwidth}\centering
    \resizebox{1.\textwidth}{!}{
	\begin{tabular}{>{\centering}m{10pt}|>{\centering}m{10pt}|>{\centering}m{10pt}|>{\centering}m{10pt}|>{\centering}m{10pt}|>{\centering}m{10pt}|>{\centering}m{10pt}|>{\centering}m{10pt}|>{\centering}m{10pt}|c|}
   \cellcolor{newgray}  & \cellcolor{newgray} $\Delta$ & \cellcolor{newgray} $\varphi$ & \cellcolor{newgray} $N$ & \cellcolor{newgray} $\Sigma$ & \cellcolor{newgray} $Q_7$ & \cellcolor{newgray} $T_1^\dagger$ & \cellcolor{newgray} $U_1$ & \cellcolor{newgray} $\bar{V}_2^\dagger$ & \cellcolor{newgray} $U_3$ \\
    \hline
    \cellcolor{newgray} $\Delta$ &  & {\small \CIRCLE} &  &  & {\small \CIRCLE} & {\small \CIRCLE} &  &  &  \\
    \hline
    \cellcolor{newgray} $\varphi$ & {\small \CIRCLE} &  & ${\small \Circle}$ & ${\small \Circle}$ &  &  &  &  &  \\
    \hline
    \cellcolor{newgray} $N$ &  & ${\small \Circle}$ &  &  &  &  & ${\small \Circle}$ & ${\small \Circle}$ &  \\
    \hline
    \cellcolor{newgray} $\Sigma$ &  & ${\small \Circle}$ &  &  &  &  &  & ${\small \Circle}$ & ${\small \Circle}$ \\
    \hline
    \cellcolor{newgray} $Q_7$ & {\small \CIRCLE} &  &  &  &  &  & {\small II} &  & {\small II} \\
    \hline
    \cellcolor{newgray} $T_1^\dagger$ & {\small \CIRCLE} &  &  &  &  &  &  & {\small II} &  \\
    \hline
    \cellcolor{newgray} $U_1$ &  &  & ${\small \Circle}$ &  & {\small II} &  &  & {\small I} &  \\
    \hline
    \cellcolor{newgray} $\bar{V}_2^\dagger$ &  &  & ${\small \Circle}$ & ${\small \Circle}$ &  & {\small II} & {\small I} &  & {\small I} \\
    \hline
    \cellcolor{newgray} $U_3$ &  &  &  & ${\small \Circle}$ & {\small II} &  &  & {\small I} &  \\
    \hline
\end{tabular}

    }
     \end{minipage}
     \end{minipage}
  \end{tabular}
        \caption{Tree-level UV-completions of dimension-7 $\Delta L=2$ operators. A Roman number denotes the topology, where an asterisk indicates a two-loop diagram obtained by an additional $W$-boson loop. An empty (filled) circle indicates the Weinberg operator (radiative correction to the type-II seesaw diagram).}
       \label{tab:tree_ops}
    \end{table}

In Tab.~\ref{tab:tree_ops}, the combinations of BSM fields that lead to tree-level UV-completions of the $\Delta L=2$ dimension-7 SMEFT operators are shown for one operator at a time, along with the corresponding radiative neutrino mass topology. A Roman number in a given entry of these tables indicates that the corresponding neutrino mass topology shown in Fig.~\ref{fig:topsIandII} or~\ref{fig:topIII} is generated by the combination of the two fields at the left and top that have an overlapping row and column at this entry. Entries with an asterisk indicate that a two-loop diagram can be generated using the one-loop topology with the addition of an $H$-loop or a $W$-loop, such that the neutrino mass is generated at 2-loop order. An empty circle denotes that the dimension-5 operator is generated via the type-I or type-III seesaw, and a filled circle indicates that the corresponding neutrino mass diagram is topologically equivalent to a loop correction to the type-II seesaw diagram. 

\section{Loop neutrino masses: cut-off EFT estimates vs. exact UV results} \label{sec:neutrinomasses7}

\subsection{Single-scale EFT estimate based on cutoff regularisation}
Generally, any $\Delta L = 2$ SMEFT operator leads to a Majorana contribution to neutrino masses (even if small, e.g. in the scenarios generating pseudo-Dirac neutrino masses) at some order in perturbation theory. For dimension-7 operators with four fermions to generate neutrino masses, two fermion legs have to be closed into a loop, leading to an additional Yukawa coupling. For operators with external lepton singlets, an additional charged Higgs interaction may be needed to convert it into a lepton doublet, and external charged leptons belonging to a doublet can be converted to neutrinos via additional $W$ interactions. In Refs.~\cite{Babu:2001ex,deGouvea:2007qla,Deppisch:2017ecm} and related literature, the contributions of higher-dimensional effective interactions to neutrino mass via such loop diagrams have been estimated using a cut-off approximation, in which the theory is characterised by a single cut-off scale $\Lambda$. In this approach, momentum integrals are then limited by this cut-off, with NP emerging above $\Lambda$ to regularise the theory. Power counting of momentum factors can determine these divergences, but loop integrals can become complex, particularly due to momentum factors in Dirac propagators, which must appear in pairs to contribute to UV divergences. As stated in Ref.~\cite{deGouvea:2007qla}, when adding loops to induce power-law divergences proportional to $\Lambda$, the process should be limited to the point where the suppression of the induced term is no worse than $1/\Lambda$. Beyond this, additional divergences only provide small finite corrections that renormalise lower-order terms.

As further stated in Ref.~\cite{deGouvea:2007qla}, each loop comes with a numerical suppression factor due to the normalisation of the loop four-momentum integrals, specifically a factor of $1/(16\pi^2)$. This suppression offsets enhancements from divergence factors. Quadratically divergent loop diagrams are often proportional to the lowest-order contribution, multiplied by factors depending on the number of loops and $\Lambda/v$, where $v$ is the Higgs vev. For diagrams to dominate over leading-order contributions, $\Lambda$ must be generally greater than about 2 TeV (for a discussion see Ref.~\cite{Deppisch:2017ecm}). However, many loop diagrams are logarithmically divergent or even convergent, enhancing LNV rates inefficiently at the low scales accessible to future experiments. \\
With all these considerations and caveats, the expressions for neutrino masses shown in the third column of Tab.~\ref{tab:dim7opExp} can be obtained. Below, we will present an alternative to this prescription that, as will be argued, better matches the results stemming from the UV-complete scenarios.
\setlength{\extrarowheight}{4pt}
\begin{table}[t]
	\centering
	\begin{tabular}{c >{\raggedright}m{50pt} >{\raggedright}m{160pt} >{\raggedright}m{120pt} c }
		& $\mathcal{O}$ & Operator & EFT Neutrino mass  & \\[2pt]
		\cmidrule[1pt]{2-4}
		& $\mathcal{O}_{LH}^{pr}$ & $\epsilon_{ij}\epsilon_{mn}\lrs{\overline{L_p^c}{}^iL_r^m}H^jH^n\lrs{H^\dagger H}$ & $\big(\frac{1}{16\pi^2}+\frac{v^2}{\Lambda^2}\big)\frac{v^2}{\Lambda}$  & \\[2pt]
		\cline{2-4}
		& $\mathcal{O}_{LeHD}^{pr}$ & $\epsilon_{ij}\epsilon_{mn}\lrs{\overline{L_p^{c}}{}^i\gamma_\mu e_r}H^j\lrs{H^miD^\mu H^n}$ & $y_e^{ri}\frac{g'}{(16\pi^2)^2}\frac{v^2}{\Lambda}$  & \\[2pt]
		\cline{2-4}
		& $\mathcal{O}_{\bar{e}LLLH}^{prst}$ & $\epsilon_{ij}\epsilon_{mn}\lrs{\overline{e_p} L_r^i}\lrs{\overline{L_s^c}{}^j L_t^m}H^n$ & $\lrs{y_e^{pr} +y_e^{ps} +y_e^{pt}}\frac{1}{16\pi^2}\frac{v^2}{\Lambda}$  &   \\[2pt]   
		\cline{2-4}
		& $\mathcal{O}_{\bar{d}LueH}^{prst}$ & $\epsilon_{ij}\lrs{\overline{d_p}L_r^i}\lrs{\overline{u_s^c}e_t}H^j$ & $\sum_{ij}y_u^{sj}y_d^{jp}y_e^{ti}\frac{1}{(16\pi^2)^2}\frac{v^2}{\Lambda}$  &   \\[2pt]
		\cline{2-4}
		& $\mathcal{O}_{\bar{d}LQLH1}^{prst}$ & $\epsilon_{ij}\epsilon_{mn}\lrs{\overline{d_p}L_r^i}\lrs{\overline{Q_s^c}{}^jL_t^m}H^n$ & $y_d^{ps}\frac{1}{16\pi^2}\frac{ v^2}{\Lambda}$  &  \\[2pt]
		\cline{2-4}
		& $\mathcal{O}_{\bar{d}LQLH2}^{prst}$ & $\epsilon_{im}\epsilon_{jn}\lrs{\overline{d_p}L_r^i}\lrs{\overline{Q_s^c}{}^jL_t^m}H^n$ & $y_d^{ps} \frac{g^2}{(16\pi^2)^2}\frac{v^2}{\Lambda}$  &   \\[2pt]
		\cline{2-4}
		& $\mathcal{O}_{\bar{Q}uLLH}^{prst}$ & $\epsilon_{ij}\lrs{\overline{Q_p}u_r}\lrs{\overline{L_s^c}L_t^i}H^j$ & $y_u^{pr}\frac{1}{16\pi^2}\frac{ v^2}{\Lambda}$  & \\[2pt]
	\end{tabular}
	\caption{List of the $\Delta L = 2$ dimension-7 operators that appear in Tabs.~\cref{tab:tree_ops}, along with the cut-off expression for the neutrino mass~\cite{deGouvea:2007qla}.}
	\label{tab:dim7opExp}
\end{table}
%

%%%%%%%%%%%%%%%%%%%%%%%%%%%%%%%%%%%%%%
\subsection{An explicit UV model example and exact loop computation}
\label{sec:leptoquark}
%%%%%%%%%%%%%%%%%%%%%%%%%%%%%%%%%%%%%%%%%%%%%%%%%%%%%%%%%%%%%%%%%%%%%%%%%%%%%%%%%

In this section, we provide a more in-depth analysis of one of the simplified models as an example. This model consists of an extension to the SM with two scalar leptoquarks, $\tilde R_2$ and $S_1$. The Lagrangian is given by~\cite{AristizabalSierra:2007nf,Dorsner:2017wwn,Cata:2019wbu}
\begin{equation}
	\label{eq:fullLQlagrangian}
	\begin{aligned}
		\mathcal{L} &= \mathcal{L}_{\text{SM}} - \tilde{R}_2^{\dagger}(\Box + m_{\tilde{R}_2}^2)\tilde{R}_{2} - S_1^*(\Box + m_{S_1}^2)S_1 + \mu S_1 H^{\dagger}\tilde{R}_{2} \\
		&- g_1^{ij}\bar{L}_{i} i\sigma_2\tilde{R}_{2}^{*}\overline{d}^c_{j} - g_2^{ij}Q_i\epsilon L_jS_1 -g_3^{ij}\bar{u}_{i}^ce_jS_1
  + \text{h.c.}\, ,
	\end{aligned}
\end{equation}
where $\Box=\eta^{\mu\nu}\partial_\mu\partial_\nu$ is the d'Alembert operator, $i$, $j$, $k$ are flavour indices, and $\epsilon$ denotes that the $SU(2)_L$ indices of the two doublets of this term should be contracted via the antisymmetric Levi-Civita tensor. If both $\tilde R_2$ and $S_1$ are heavy with respect to the electroweak symmetry breaking scale, Eq.~\eqref{eq:fullLQlagrangian} leads to the generation of operator $\mathcal{O}_{\bar dLQLH1}$ and $\mathcal{O}_{\bar dLueH}$ 
at the low scale, with corresponding Wilson coefficients
\begin{equation}
	\label{eq:thefinalkaonscale}
	C^{ijkn}_{\bar dLQLH1} = -\frac{\mu g_{1}^{ni} g_{2}^{kj}}{m_{\tilde{R}_2}^2 m_{S_1}^2}\, , \qquad C^{ijkn}_{\bar dLueH} = \frac{\mu g_{1}^{ji} g_3^{kn}}{m_{\tilde{R}_2}^2 m_{S_1}^2}\, .
\end{equation}
\begin{figure}[t]
	\centering
 \includegraphics[width=0.40\textwidth]{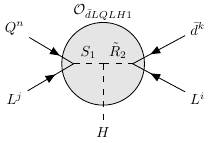}
	\includegraphics[width=0.54\textwidth]{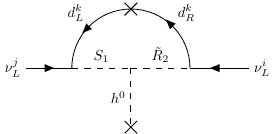}\\
 \includegraphics[width=0.40\textwidth]{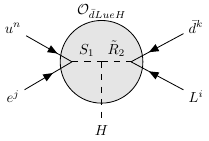}
	\includegraphics[width=0.54\textwidth]{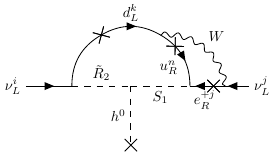}
	\caption{\textbf{Left column: }Operators $\mathcal{O}_{\bar dLQLH1}$ (top) and $\mathcal{O}_{\bar dLueH}$ (bottom) generated by the model described in the text. \textbf{Right column:} Radiative neutrino mass diagrams generated in the model described in the text, corresponding to the set of coupling constants that lead to $\mathcal{O}_{\bar dLQLH1}$ (top) and $\mathcal{O}_{\bar dLueH}$ (bottom).
 }
	\label{fig:mnudiagram2}
\end{figure}
In Fig.~\ref{fig:mnudiagram2} (left column), we show how these operators are generated in this model. This operator can then lead to \ovbb decay and LNV rare kaon decays~\cite{DeRomeri:2023cjt,Fajfer:2024uut,Dev:2024tto}.

This model also leads to radiative neutrino masses~\cite{Chua:1999si,Mahanta:1999xd,AristizabalSierra:2007nf,Dorsner:2017wwn,Cata:2019wbu,Babu:2019mfe,Deppisch:2020oyx,Fajfer:2024uut,Dev:2024tto,Fridell:2024otl} as well as leptogenesis~\cite{Fridell:2024otl}. At lowest order, we have a 1-loop neutrino mass shown in Fig.~\ref{fig:mnudiagram2} (top right). If the coupling $g_2$ is suppressed, neutrino masses can still be generated at 2-loop order via $g_3$ as shown in Fig.~\ref{fig:mnudiagram2} (bottom right), corresponding to the set of couplings that generate $\mathcal{O}_{\bar dLueH}$. 
The mass matrix for the leptoquark pair is given by
\begin{equation}
	\label{eq:LQmassmat}
	M^2=\begin{pmatrix}
		m_{\tilde{R}_2}^2 & \mu v\\
		\mu v & m_{S_1}^2
	\end{pmatrix}\, .
\end{equation}
Diagonalising this matrix leads to two mass eigenstates $m_{\text{LQ}_{1,2}}$ given by
\begin{equation}
	m_{\text{LQ}_{1,2}}^2=\frac{1}{2}\lr{m_{\tilde R_2}^2+m_{S_1}^2\mp \sqrt{(m_{\tilde R_2}^2-m_{S_1}^2)^2+\mu^2v^2}}\, .
\end{equation}
Using these mass eigenstates, we can express the 1-loop neutrino mass from Fig.~\ref{fig:mnudiagram2} (top right) as~\cite{Babu:2019mfe}
\begin{equation}
	\label{eq:LQ1loopmass}
	(m_{\nu})_{ij}=\sum_{k}\frac{3\sin\lr{2\theta}y_{kk}^dv{g}_1^{ik}{g}_2^{kj}}{32\pi^2}\log\frac{m_{\text{LQ}_2}^2}{m_{\text{LQ}_1}^2}\, ,
\end{equation}
where $y^d$ is the SM down-type quark Yukawa coupling matrix. The mixing angle $\theta$ is given by~\cite{AristizabalSierra:2007nf,Babu:2010vp,Dorsner:2017wwn}
\begin{equation}
	\tan(2\theta)=\frac{2\mu v}{m_{\tilde{R}_2}^2-m_{S_1}^2}\, ,
\end{equation}
where $V$ is the CKM matrix.  
The 2-loop radiative neutrino mass shown in Fig.~\ref{fig:mnudiagram2} (bottom right) can be expressed as~\cite{Babu:2010vp}
 \begin{eqnarray}
 \label{eq:thefinalnumass}
 (m_\nu)_{ij} = &&\frac{3\sin(2\theta)    v^3g^2}
 {(16\pi^2)^2 \, m_{\text{LQ}_1}^2} \left( g_1^{ik} y^d_{kk} V^T_{kl} y^u_{ll} (g_2)^{lj} y^e_{jj} + y^e_{ii} (g_2^T)^{il} y^u_{ll} V_{lk} y^d_{kk} (g_1^T)^{kj} \right)\nonumber\\
&&\times I_{jkl}(m_{\text{LQ}_1}^2,m_{\text{LQ}_2}^2,m_W^2)\, ,
 \end{eqnarray}
 where for massless SM fermions, the function $I_{jkl}(m_{\text{LQ}_1}^2,m_{\text{LQ}_2}^2,m_W^2)$ simplifies to~\cite{Babu:2010vp}
 \begin{equation}
 	\label{eq:twoloopfunction}
 	\begin{aligned}
 		I(m_{\text{LQ}_1}^2,m_{\text{LQ}_2}^2,m_W^2) &\approx \lr{1-\frac{m_{\text{LQ}_1}^2}{m_{\text{LQ}_2}^2}}\times\\
 		&\Bigg[1+\frac{\pi^2}{3}+\frac{m_{\text{LQ}_1}^2\log\frac{m_{\text{LQ}_2}^2}{m_W^2}-m_{\text{LQ}_2}^2\log\frac{m_{\text{LQ}_1}^2}{m_W^2}}{m_{\text{LQ}_1}^2-m_{\text{LQ}_2}^2}+\\
 		&\quad \frac{1}{2}\frac{m_{\text{LQ}_1}^2\log^2\frac{m_{\text{LQ}_2}^2}{m_W^2}-m_{\text{LQ}_2}^2\log^2\frac{m_{\text{LQ}_1}^2}{m_W^2}}{m_{\text{LQ}_1}^2-m_{\text{LQ}_2}^2}\Bigg]\, .
 	\end{aligned}
 \end{equation}
Below, we compare the 1- and 2-loop 
neutrino mass expressions in this model, given by Eqs.~\eqref{eq:LQ1loopmass} and \eqref{eq:thefinalnumass}, with the cut-off-regularisation-based estimates in Tab.~\ref{tab:dim7opExp}. The cut-off-regularisation-based one-loop neutrino mass estimate for $\mathcal{O}_{\bar dLQLH1}$ in the EFT shown in Tab.~\ref{tab:dim7opExp} can be expressed in the flavour basis as
\begin{equation}
	\label{eq:LQEFTmass}
	(m_\nu)_{ij} =\frac{y^d_{nk}}{16\pi^2}\frac{v^2}{\left(C^{kinj}_{\bar dLQLH1}\right)^{-1/3}}\, ,
\end{equation}
which should be compared against the exact model result Eq.~\eqref{eq:LQ1loopmass}. Similarly, the two-loop neutrino mass estimate for $\mathcal{O}_{\bar{d}LueH}$ in the cut-off regularisation can be expressed as

\begin{equation}
	\label{eq:LQEFTmass-2loop}
	(m_\nu)_{ij}=
g^2 y^u_{sl}V_{lr}y^d_{rp}y^e_{jk}\frac{1}{(16\pi^2)^2}\frac{v^2}{\left(C_{\bar{d}LueH}^{pisk}\right)^{-1/3}}\, ,
\end{equation}
 which should be compared against the exact model result Eq.~\eqref{eq:thefinalnumass}.

\begin{figure}
	\centering	
	\includegraphics[width=0.49\textwidth]{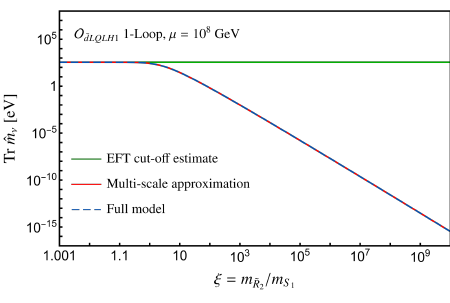} \hfill
	\includegraphics[width=0.49\textwidth]{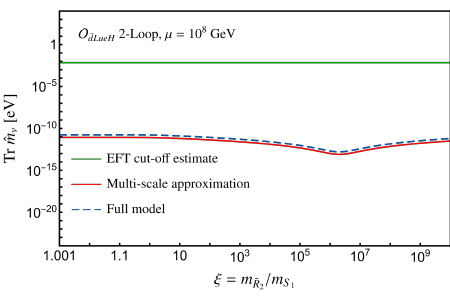}\\
	\caption{The sum of neutrino masses as a function of the degree of hierarchy between $m_{\tilde R_2}$ and $m_{S_1}$, assuming $m_{\tilde R_2}>m_{S_1}$, as quantified by the parameter $\xi$, at 1-loop (left) and 2-loop (right) using $\mu=10^8$~GeV.
 The green line shows the cut-off estimate from Eq.~\eqref{eq:LQEFTmass}, the blue line shows the full model-based expression from Eqs.~\eqref{eq:LQ1loopmass} (1-loop) and~\eqref{eq:thefinalnumass} (2-loop). 
 The red line shows the expression obtained in the simplified model estimate.}
	\label{fig:comparemnu}
\end{figure}

Comparing the exact loop results with the cut-off regularisation, we notice that the exact results are sensitive to the hierarchy of scales between the two heavy fields. We can vary the ratio between $m_{\tilde R_2}$ and $m_{S_1}$ to show the dependence of the neutrino mass on the hierarchy of scales between the two heavy fields. To do this, we define the hierarchy parameter $\xi$ as
\begin{equation}
	\xi \equiv \frac{\text{max}\left(m_{\Phi_1},m_{\Phi_2}\right)}{\text{min}\left(m_{\Phi_1},m_{\Phi_2}\right)}\, ,
\end{equation}
where in this model we have $\Phi_1\to \tilde R_2$ and $\Phi_2\to S_1$. In Fig.~\ref{fig:comparemnu} we compare the different neutrino mass expressions for the benchmark points described above with $m_{\tilde R_2} > m_{S_1}$ at both 1- and 2-loop.
We assume normal ordering and $CP$ conservation in the neutrino sector, and use the neutrino mixing angles given by \textsc{NuFIT v6.0}~\cite{Esteban:2024eli}, explicitly $\theta_{12}=33.68^\circ,$ $\theta_{23}=44.3^\circ,$ and $\theta_{13}=8.56^\circ$.

For the 1-loop mass, we see that the expressions agree quite well for small $\xi$, i.e.\ when there is a small mass difference between $m_{\tilde R_2}$ and $m_{S_1}$. For larger $\xi$, the neutrino mass is smaller in the exact model-based expression, while remaining constant in the cut-off estimate, leading to an over-estimation of the constraint on the Wilson coefficient coming from neutrino masses. Compared to the leptoquark model expressions, we see that capturing the effect of a sizable hierarchy in the internal degrees of freedom of an operator is crucial for correctly determining the available region of parameter space. 

On the other hand, for the 2-loop case, we notice that the estimate for the neutrino mass based on cut-off regularisation using the EFT operator, as given in the third column of Tab.~\ref{tab:dim7opExp}, fails to reproduce the correct order of magnitude when compared to the exact results, even when the heavy new physics degrees of freedom are non-hierarchical.

In the following section, we will discuss this failure of the cut-off regularisation based estimates and the motivation to seek a new approach.

 %%%%%%%%%%%%%%%%%%%%%%%%%%%%%%%%%%%%%%%%%%%%%%%%%%%%%%%%%%%%%%%
 %%%%%%%%%%%%%%%%%%%%%%%%%%%%%%%%%%%%%%
 \subsection{Failure of the cut-off regularisation and motivations to go beyond}
 \label{sec:doomed_cut-off}
Estimates of the neutrino mass based on the cut-off method supplemented by naive dimensional analysis, e.g. the expressions given in the third column of Tab.~\ref{tab:dim7opExp}, have been widely used in the literature to estimate the constraint of small neutrino masses on higher-dimensional $\Delta L = 2$ operators. From the explicitly worked out examples in the previous subsections and particularly as visible in Fig.~\ref{fig:comparemnu}, it should be apparent that this treatment not only fails to capture the effect of a non-vanishing hierarchy in the heavy internal degrees of freedom in the 1-loop case, but also gives an estimate for the neutrino mass in the 2-loop case that is by ten orders of magnitude larger than in the UV model. The underlying reasons for these two observations are, however, different in their nature, and we want to discuss them in the following.

From an EFT point of view, an insertion of a higher dimensional operator with dimension $\mathcal{D}$ inside a diagram corresponding to a kinematic energy scale $p$ should give a contribution to the amplitude in $d$ space-time dimensions of the order $\mathcal{A}\propto (p/\Lambda)^{\mathcal{D}-d}$, where $\Lambda$ is the heavy scale suppressing the higher dimensional operator. This is often called the EFT power counting formula~\cite{Manohar:2018aog}. In the case of loop diagrams, the loop momentum $k_l$ has to be integrated over all values $-\infty \leq k_l\leq +\infty$ and consequently the EFT expansion in $k_l/\Lambda$ breaks down; however, the power counting formula is known to remain valid. This can be seen by comparing the loop integrals with the method of residues for evaluating complex integrals, as follows. Consider a general loop integral in the EFT (after already integrating out the heavy scales) of the form
 \begin{equation}
 	I_L= \int_{k_{l}}  d^dk_{l} \frac{(k_{l}^2)^{a}}{(k_{l}^2-m^2)^{b}}\,.
 \end{equation}
We notice that for a specified $d$, $a$ and $b$, a contour can be chosen such that the integral vanishes sufficiently fast as $k_l\rightarrow \infty$, and the contour at infinity gives a vanishing contribution. The integrand is then given by the sum of the residues at the poles given by the denominators of the integrand. The poles can only be a function of the masses of light degrees of freedom in the EFT and the external momenta. Consequently, there are no compensating factors of $\Lambda$ in the numerator, and $\Lambda$ should appear only due to the EFT vertex factor. 

The discrepancy that we observe for large hierarchies in the 1-loop example, see Fig.~\ref{fig:comparemnu} (left), is caused by the missing hard region of the loop integral. The naive EFT estimate only captures the soft region of the loop integral $m\sim p\sim k_l \ll M$, where $m$ is the mass of the light degrees of freedom, $p$ is a linear combination of the external momenta, $k_l$ is the loop momenta, and $M$ is the heavy mass-scale in the EFT. The missing piece is the hard region of the loop integral $m\sim p\ll k_l \sim M$, which can contain non-local dependence on the heavy mass scale $M$ (powers of $M$ in the numerator). Consequently, it can modify the behaviour obtained within the EFT, which is valid only in the soft region. This is precisely the case in the 1-loop case in the example of the previous subsection, where the power-counting estimate within the EFT, as discussed above in the one-loop case, needs to be supplemented by the hard region to obtain the correct behaviour as seen in Eq.~\eqref{eq:LQ1loopmass}. Only then can hierarchies of new physics be consistently taken into account and lead to a reliable estimate, as demonstrated by the multi-scale approximation in Fig.~\ref{fig:comparemnu} (left).

Following the above simple argument, the two-loop diagram in the EFT corresponding to the bottom diagram of Fig.~\ref{fig:mnudiagram2}, after integrating out the heavy degrees of freedom, should scale as $I_{l}\propto v^4/\Lambda^3$. However, it can be seen from the estimate in Tab.~\ref{tab:dim7opExp} that the corresponding estimate using the cut-off fails to reproduce the correct power counting behaviour. This is because this naive estimate corresponds to a diagram similar to the bottom-right diagram of Fig.~\ref{fig:mnudiagram2}, with the $W$ boson replaced by a Higgs boson, and such a diagram requires only one SM fermion mass-insertion (denoted by crosses in the diagram) either in the up or down quark line. In the SM, the up and down couplings to the Higgs boson come with a relative negative sign. Therefore, when the diagrams with mass insertion in the up-type quark and the down-type quark lines are added up, the whole contribution due to the charged Higgs loop vanishes, see e.g.~\cite{Babu:2010vp} for a discussion. The next leading contribution comes from the diagram in the bottom-right of Fig.~\ref{fig:mnudiagram2}, which clearly should have a different EFT power-counting behaviour due to a different number of mass insertions required. We have checked explicitly that in the two-loop diagram example in the previous subsection, corresponding to the operator $\mathcal{O}_{\bar dLQLH2}$ and the bottom diagram in Fig. 3, can be much better estimated by
 \begin{equation}
 	y_e y_d y_u \frac{g^2}{(16\pi^2)^2}\frac{v^4}{\Lambda^3} \, .
 \end{equation}
which follows from the simple power counting argument. 

The EFT power counting argument holds in any loop order. However, if one works with in the above way with just the EFT contact interaction after integrating out the heavy degrees of freedom, the corresponding estimate only captures the soft region of the loop integral $m\sim p\sim k_l \ll M$, where $m$ is the mass of the light degrees of freedom, $p$ is a linear combination of the the external momenta, $k_l$ is the loop momenta, and $M$ is the heavy mass-scale in the EFT. The missing piece is the hard region of the loop integral $m\sim p\ll k_l \sim M$, which can contain non-local dependence on the heavy mass scale $M$ (powers of $M$ in the numerator). Consequently, it can modify the simple and correct power-counting behaviour obtained within the EFT, valid only in the soft region. This is precisely the case in the one-loop case in the example of the previous subsection, where the power-counting estimate within the EFT, as discussed above, needs to be supplemented by the hard region to obtain the correct behaviour as seen in Eq.~\eqref{eq:LQ1loopmass}. The estimates in Tab.~\ref{tab:dim7opExp} not only use a native cut-off scheme but also employ a number of ``dimensional analysis" rules based on subtle generalisations of dimensional regularisation based results. Consequently, they seem to work fairly well in most cases. However, in certain examples, such estimates can break down because of physical cancellations of the leading diagrams and in almost all examples, will deviate from the exact result for a hierarchy between the heavy scales.  
 
It is well known that the dimensional regularisation provides a consistent way to deal with the loops consistently with the correct power counting behaviour. In a fully consistent top-down EFT approach, a complete EFT basis with renormalisation group (RG) evolution needs to be supplemented by the matching of the UV model to EFT to capture the non-local effects. Furthermore, for hierarchy between the heavy new physics fields, the field needs to be integrated out one at a time with subsequent matching and running. In the following section, we proceed to discuss a relatively simpler approach to estimating loop neutrino masses using the dimensional regularisation approach based on the method of regions~\cite{Manohar:2018aog}, which not only ensures a correct power counting behaviour, but also provides a handle on the hierarchy between the heavy new physics fields.
 %	
 %%%%%%%%%%%%%%%%%%%%%%%%%%%%%%%%%%%%%%%%%%%%%%%%%%%%%%%%%%%%%%%%%%%%%%%%%%%%%%%%%

%%%%%%%%%%%%%%%%%%%%%%%%%%%%%%%%%%%%%%%%%%%%%%%%%%%%%%%%%%%%%%%%%%%%%%%%%%%%%%%%%
\section{A simplified multi-scale approach to loop neutrino masses}
\label{sec:dim_reg}
It is well known that a much more robust approach compared to that based on cut-off regularisation is provided by the renormalisation group (RG) improved perturbation theory~\cite{Manohar:2018aog}. However, the derivation and implementation of the full RG equations order by order for an independent basis of higher dimensional operators is often quite cumbersome. For instance, recently in Refs .~\cite {Liao:2019tep,Zhang:2023kvw,Zhang:2023ndw} the RG equations for dimension-7 LNV operators at one-loop order have been derived. In Ref.~\cite{Chala:2021juk} a leading-log approximation for light neutrino masses arising from LNV operators up to dimension-7 was presented, keeping only the largest SM Yukawa couplings. A detailed technical treatment of the full RG equations is beyond the scope of this work. Instead, in what follows, we will discuss the limitations of the cut-off-based approach, which leads to a poor approximation when the heavy new physics degrees of freedom generating radiative neutrino masses are hierarchical in mass. We will show that a simple dimensional regularisation based approach, combined with simple scaling arguments, can approximate the loop neutrino masses arising from higher dimensional operators much more accurately compared to the cut-off based approach. This simplified multi-scale formalism gives a quick way to accurately estimate the loop neutrino masses using simple momentum expansion arguments without a full implementation of the full set of RGE equations, which can often be a very cumbersome exercise. 

To highlight the limitation of the cut-off regularisation approach, let us consider the simple toy Lagrangian,
\begin{align} \label{full-theory}
	\mathcal{L}^{\text{full}} =& \int d^4 x \left\{ - \frac{1}{2} \partial_\mu \Phi \, \partial^\mu \Phi - \frac{1}{2} M^2 \Phi^2 + \bar \psi (i \gamma^\mu \partial_\mu - m ) \psi +  i Y\, \Phi \bar  \psi \gamma^5 \psi - \frac{1}{4!} \lambda \Phi^4 \right\} \, ,
\end{align}
where $\Phi$ with mass $M$ is a heavy pseudo-scalar and $\psi$ is a light fermion with mass $m$, such that $M\gg m$. If the typical energy scale of an observable $E$ is such that $m\leq E \ll M$, then a two-point one-loop fermionic tadpole correction in the EFT obtained after integrating out the heavy scalar $\Phi$ is given by 
\begin{align} \label{eq:EFT2}
	G_2^{\text{EFT}}  = - \, \frac{Y^2 m}{M^2} \int \frac{d^4 k}{(2 \pi)^4} \, \frac{1}{k^2 + m^2 - i \epsilon} = - \, \frac{ m}{\Lambda^2} \int \frac{d^4 k}{(2 \pi)^4} \, \frac{1}{k^2 + m^2 - i \epsilon}  \, ,
\end{align}
where to write the last equality, we have used the definition that the EFT scale $\Lambda\equiv M/Y$. According to the cut-off-regularisation-based approach mentioned above, one would cut off the integral over $k$ at the heavy scale $\Lambda$, since the EFT obtained by integrating out $\Phi$ is expected to be valid at energies $E < \Lambda$~\cite{Wilson:1983xri}. This leads to a resulting factor of $\Lambda^2$ in the numerator, which would cancel against the $1/\Lambda^2$ in the Wilson coefficient, giving a result that is independent of the heavy UV scale $\Lambda$. This is not consistent with the power counting rule, suggesting a dependence $E^2 / \Lambda^2$.

Using a dimensional regularisation approach instead, the integral is generalised to $d = 4 -\varepsilon$ dimensions, where $\varepsilon$ denotes the deviation from $d=4$. Then the integral in Eq. \eqref{eq:EFT2} can be expressed in the form
\begin{align} \label{eq:EFT3}
	G_{2, \text{dim-reg}}^{\text{EFT}}  =- \, \frac{ \tilde{\mu}_r^\varepsilon m}{\Lambda^2} \int \frac{d^{4-\varepsilon} k}{(2 \pi)^{4-\varepsilon}} \, \frac{1}{k^2 + m^2 - i \epsilon} =\frac{i}{16 \pi^2} \frac{m^3}{\Lambda^2} \left[ \frac{2}{\varepsilon} + 1 + \log\left(\frac{\mu_r^2}{m^2}\right) + \mathcal{O}(\varepsilon) \right] ,
\end{align}
where the renormalisation scale $\tilde \mu_r$ replaces $Y \to Y \, \tilde{\mu}_r^{\varepsilon/2}$ to ensure that $Y$ is dimensionless, and where $\mu_r$ is given by $\mu_r = \sqrt{4 \pi} e^{-\gamma} \tilde \mu_r$, with $\gamma$ being the Euler constant. The UV-divergent piece proportional to $1/\varepsilon$ drops out in the $\overline{\text{MS}}$-scheme, and we obtain a loop integral result which is now consistent with the power-counting expectation.

While the EFT result in Eq.~\eqref{eq:EFT3} can be recovered by evaluating the integral in the full theory, c.f.\ Eq.~\eqref{full-theory}, using dimensional regularisation and then expanding in the heavy mass scale $M$, it is well known that the loop integral and the series expansion do not commute in general~\cite{Manohar:2018aog}. This leads to a difference in the exact theory vs. the EFT estimation of loops. This difference is analytic in the IR scale $m$ and leads to a matching contribution, which needs to be added to the EFT result to correctly reproduce the full theory result. To be more concrete, the full theory computation differs from the result in Eq.~\eqref{eq:EFT3} by terms proportional to $\log(M^2/\mu_r^2)$ at one-loop order and by higher powers of the same $\log$ for higher order loops~\cite{Manohar:2018aog}. 

A particularly interesting scenario is when two largely different heavy NP scales are present in the full theory $M\gg m\gg E$, where $E$ is the energy scale of the experimental observable we are interested in. In the presence of a large hierarchy between different scales $M$ and $m$ in the full theory, the relatively large logs appearing in the loop integrals of the form $\log(m^2/M^2)$ can be split into two contributions of the form $\log(M^2/m^2)=\log(M^2/\mu_r^2)-\log(m^2/\mu_r^2)$. The first part can be absorbed in the Wilson coefficient (upon integrating out the large scale $M$) by choosing the renormalisation scale $\mu_r=M$ and doing the matching procedure at $\mu_r=M$. The second part, on the other hand, can be obtained by calculating the one-loop corrections in the EFT at $\mu_r=m$, which can be done using the RG equations capturing the running of the Wilson coefficients from $\mu_r=M$ to $\mu_r=m$. We note that the EFT after integrating out $M$ depends on either ratios of the form $(m^2/M^2)$ or $\log(M^2/\mu_r^2)$. If we evaluate the Wilson coefficient at $\mu_r=M$, then the terms proportional to $\log(M^2/\mu_r^2)$ vanish and only terms proportional to powers of $(m^2/M^2)$ survive. 

\begin{figure}
	\centering
\includegraphics[scale=1.7]{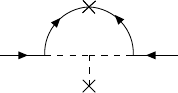} \raisebox{9.3mm}{\LARGE $\,\Rightarrow\,$}
\includegraphics[scale=1.7]{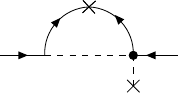} \raisebox{9.5mm}{\LARGE $\, +\,$}
\raisebox{9.5mm}{\includegraphics[scale=1.7]{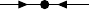}}
\caption{One-loop neutrino mass diagrams in the full and effective theories. The crosses represent fermion mass insertions, and the thick circle represents contact interactions.}
\label{fig:model+eft}
 \end{figure}

\subsection{Simplified multi-scale formalism for one-loop neutrino masses}
With the above prescription in mind, let us now consider the one-loop neutrino mass in the model example discussed in section~\ref{sec:leptoquark}, c.f.\ the top diagram in Fig.~\ref{fig:mnudiagram2}. We are interested in the case where the heavy new physics is hierarchical in mass. Let us consider for example the scenario where $m_{\tilde{R}_2} \gg m_{S_1}$. The one-loop neutrino mass diagram in the UV model and how it matches into the EFT obtained after integrating out the heaviest degree of freedom $\tilde{R}_2$ are shown in Fig.~\ref{fig:model+eft}, where the last diagram to the right symbolises any potential additional effective contributions arising from integrating out the heaviest degree of freedom. 

In the full theory, the one-loop neutrino mass diagram in dimensional regularisation gives the integral
\begin{equation}
  I_1 = c_m \, \mu_r^{2 \varepsilon}\int \frac{d^dk}{(2\pi)^d} \frac{1}{k^2}\frac{1}{k^2 - m_{S_1}^2 } \frac{1}{k^2 - m_{\tilde{R}_2}^2}\, ,
\end{equation}
where the dimensionful prefactor is given by $c_m=g_1 g_2 v \mu m_d^2$, and we neglect the light down-type quark masses in the denominators for simplicity.
Now the neutrino mass in the intermediate EFT (I-EFT) obtained after integrating out the heaviest degree of freedom $\tilde{R}_2$ can be obtained by expanding out the above integral in the heavy mass $m_{\tilde{R}_2}$
\begin{equation}
  I_1^{\text{I-EFT}} = c_m \, \mu_r^{2 \varepsilon}\int \frac{d^dk}{(2\pi)^d} \frac{1}{k^2}\frac{1}{k^2 - m_{S_1}^2 } \left[ -\frac{1}{ m_{\tilde{R}_2}^2}\left(1 +\frac{k^2}{ m_{\tilde{R}_2}^2} +\frac{k^4}{ m_{\tilde{R}_2}^4} +\cdots\right) \right]\,.
\end{equation}
In Fig.~\ref{fig:model+eft}, this corresponds to the left diagram of the EFT case, while there is no tree-level effective contribution. Integrating term by term yields
\begin{eqnarray}
    I_1^{\text{I-EFT}} = &-&c_m \, \frac{i}{16 \pi^2} \frac{1}{m_{\tilde{R}_2}^2} \Bigg[  \left(1+\log \left(\frac{\mu_r^2}{m_{S_1}^2}\right) \right) + \frac{m_{S_1}^2}{m_{\tilde{R}_2}^2}\left(1+\log \left(\frac{\mu_r^2}{m_{S_1}^2}\right) \right) \nonumber\\ && + \frac{m_{S_1}^4}{m_{\tilde{R}_2}^4}\left(1+\log \left(\frac{\mu_r^2}{m_{S_1}^2}\right) \right) +\cdots \Bigg] \nonumber\\
     \approx &-& c_m \, \frac{i}{16 \pi^2} \frac{1}{(m_{\tilde{R}_2}^2-m_{S_1}^2) } \left[ 1+ \log \left( \frac{\mu_r^2}{m_{S_1}^2}\right) \right]\, .
\end{eqnarray}

Now the analytic contributions $I_1^{\text{ana}}$ missed in the EFT expansion (containing terms proportional to $\log m_{\tilde{R}_2}$) can be obtained by performing an expansion of $I_1$ and $I^{\text{I-EFT}}_1$ in $m_{S_1}$ and taking $I^{\text{ana}}_1={I_1}_{\text{exp}}-{I_1}^{\text{I-EFT}}_{\text{exp}}$ where the subscript ``exp" denotes an expansion in the lighter scale $m_{S_1}$. The expansions ${I_1}_{\text{exp}}$ and ${I_1}^{\text{EFT}}_{\text{exp}}$ are given by
\begin{equation}
  {I_1}_{\text{exp}} = c_m \, \mu_r^{2 \varepsilon}\int \frac{d^dk}{(2\pi)^d} \frac{1}{k^2} \frac{1}{k^2 - m_{\tilde{R}_2}^2} \left[ \frac{1}{k^2}\left(1 +\frac{m_{S_1}^2}{ k^2} +\frac{m_{S_1}^4}{ k^4} +\cdots\right) \right]\, ,
\end{equation}
and
\begin{equation}
  {I_1}^{\text{I-EFT}}_{\text{exp}} = c_m \, \mu_r^{2 \varepsilon}\int \frac{d^dk}{(2\pi)^d} \frac{1}{k^2}
  \left[ \frac{1}{k^2}\left(1 +\frac{m_{S_1}}{ k^2} +\frac{m_{S_1}^2}{ k^4} +\cdots\right) \right]
  \left[ -\frac{1}{ m_{\tilde{R}_2}^2}\left(1 +\frac{k^2}{ m_{\tilde{R}_2}^2} +\cdots\right) \right]\, .
\end{equation}
Upon integrating term by term ${I_1}^{\text{I-EFT}}_{\text{exp}}$ vanishes, while $ I_{1\text{exp}}$ gives rise to a non-vanishing contribution for ${I_1}^{\text{ana}}$ 
\begin{eqnarray}
  {I_1}^{\text{ana}} &=& c_m \, \frac{i}{16 \pi^2}\left[ \frac{1}{m_{\tilde{R}_2}^2}\left(1+\log \left(\frac{\mu_r^2}{m_{\tilde{R}_2}^2}\right) \right) + \frac{m_{S_1}^2}{m_{\tilde{R}_2}^4}\left(1+\log \left(\frac{\mu_r^2}{m_{\tilde{R}_2}^2}\right) \right) +\cdots\right] \nonumber\\
  &\approx& c_m \, \frac{i}{16 \pi^2} \frac{1}{(m_{\tilde{R}_2}^2-m_{S_1}^2) } \left[ 1+ \log \left( \frac{\mu_r^2}{m_{\tilde{R}_2}^2}\right) \right]\,.
\end{eqnarray}
The whole contribution to the one-loop neutrino mass is then given by
\begin{eqnarray}
   {I_1}^{\text{total}}_{\text{MS}} &=&    {I_1}^{\text{EFT}}+{I_1}^{\text{ana}}\nonumber\\
                      &\approx&    c_m \, \frac{i}{16 \pi^2} \frac{1}{(m_{\tilde{R}_2}^2-m_{S_1}^2) }   \log \left( \frac{m_{S_1}^2}{m_{\tilde{R}_2}^2}\right)\, ,
\end{eqnarray}
where we use the subscript $\text{MS}$ to denote that this is the result using the multi-scale approach.  

The exact expression for the loop diagram in the full model, c.f.\ Eq.~\eqref{eq:LQ1loopmass}, when expanded in the heavy scale, yields the same expression. Therefore, the above prescription represents a consistent approach to estimate the loop neutrino masses starting from an EFT interaction and without the knowledge of the exact computation in the full theory. In essence, we have split the original one-loop computation involving two hierarchical new physics scales into a two-scale EFT problem. In the particular case when $m_{\tilde{R}_2} \rightarrow m_{S_1}$, the above expression in the dimensional regularisation reduces to the standard single scale EFT result using cut-off regularisation, leading to a $1/\Lambda$ dependence with $\Lambda \sim m_{\tilde{R}_2} \sim m_{S_1}$, c.f.\ Tab.~\ref{tab:dim7opExp}. 

\begin{figure}
	\centering
	\includegraphics[width=0.49\textwidth]{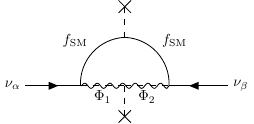}
	\includegraphics[width=0.49\textwidth]{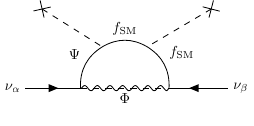}
	\caption{One-loop neutrino masses based on the simplified interactions in Eq.~\ref{eq:lagSimpMI} leading to topology I (left) and  Eq.~\ref{eq:lagSimpMII} leading to topology II.}
	\label{fig:topsIandII_sim}
\end{figure}

\subsection{Simplified multi-scale formalism for two-loop neutrino masses}
A similar prescription can also be applied to higher loop orders. However, the evaluation of the multi-loop integrals becomes quickly nontrivial, and a number of simplifying physical assumptions can help in getting relatively simpler analytic results. Let us consider the two-loop neutrino mass diagram in the model example discussed in section~\ref{sec:leptoquark}, c.f.\ the bottom diagram in Fig.~\ref{fig:mnudiagram2}. To estimate the contribution using our prescription, it suffices to approximate the $W$-boson propagator by a scalar propagator of similar mass. The complete contribution involves additional terms which are suppressed by $m_W^2/m_{\tilde{R}_2}^2$ and $m_W^2/m_{S_1}^2$, which will provide subdominant corrections. Under the above approximations, the two-loop neutrino mass diagram can be approximated as
\begin{equation}
M_\nu\simeq c'_m \int d^4 p \int d^4 q \frac{1}{p^2 } 
\frac{1}{q^2}\frac{1}{q^2-m_{W}^2}\frac{1}{(p-q)^2-m_u^2} \frac{1}{p^2-m_{S_1}^2} \frac{1}{p^2-m_{\tilde{R}_2}^2}\\ %\nonumber
\equiv 
c'_m I_2\ ,
\label{eq:2loop_approx}
\end{equation}
where $c'_m \sim \frac{N_c}{(2\pi)^8} m_d m_u m_e {(\mu v)} g^2 y_{S_1} y_{\tilde{R}_2}$, with the $y$'s denoting the relevant Yukawa interactions of the leptoquarks with the SM fermions, $N_c$ denoting the color factor, and $\mu$ denotes the trilinear scalar coupling with mass dimension one. In the above, we have ignored the flavour structure of the light new physics, which can be straightforwardly included as necessary. We have further assumed the limit of vanishing lepton- and down-type quark masses inside the loop integral. However, we note that the overall neutrino mass contribution is proportional to a prefactor containing these masses. Consequently, they cannot be taken to vanish in the full expression of neutrino mass. The integral $I$ can then be expressed in terms of the rescaled dimensionless momenta as
\begin{equation}
I_2 = \frac{1}{m_u^4} \int d^4 p \int d^4 q \frac{1}{p^2q^2}  
\frac{1}{q^2-r}\frac{1}{(p-q)^2-1} \frac{1}{p^2-t_1} \frac{1}{p^2-t_2}\ ,
\label{eq:2loop-scaled}
\end{equation}
where $r=m_W^2/m_u^2$, $t_1=m_{S_1}^2/m_u^2$ and $t_2=m_{\tilde{R}_2}^2/m_u^2$. Now, one can follow the same prescription as in the one-loop case. Assuming again that $\tilde{R}_2$ is much heavier than the momentum and $m_{S_1}$, we can expand the corresponding propagator to obtain in the I-EFT limit
\begin{equation}
I_2^{\text{I-EFT}} = \frac{1}{m_u^4} \int d^4 p \int d^4 q \frac{1}{p^2q^2}  
\frac{1}{q^2-r}\frac{1}{(p-q)^2-1} \frac{1}{p^2-t_1}  \left[ -\frac{1}{ t_2}\left(1 +\frac{p^2}{ t_2} +\frac{p^4}{{t_2}^2} +\cdots\right) \right]\ .
\label{eq:2loop-scaled1}
\end{equation}
Similarly, we can also use the same momentum expansion scheme as in the one-loop case to obtain the analytic contributions using 
\begin{equation}
{I_2}_{\text{exp}} = \frac{1}{m_u^4} \int d^4 p \int d^4 q \frac{1}{p^2q^2}  
\frac{1}{q^2-r}\frac{1}{(p-q)^2-1} 
  \left[ \frac{1}{p^2}\left(1+\frac{t_1}{ p^4} +\frac{t_1^2}{ p^4} +\cdots\right) \right] \frac{1}{p^2-t_2}
 \ ,
\label{eq:2loop-scaled2}
\end{equation}
and 
\begin{align}
{I_2}^{\text{I-EFT}}_{\text{exp}} &= \frac{1}{m_u^4} \int d^4 p \int d^4 q \frac{1}{p^2q^2}  
\frac{1}{q^2-r}\frac{1}{(p-q)^2-1}  \nonumber\\ & \quad\quad\quad\quad\quad\quad
\times\left[ \frac{1}{p^2}\left(1+\frac{t_1}{ p^2} +\frac{t_1^2}{ p^4} +\cdots\right) \right]  \left[ -\frac{1}{ t_2}\left(1 +\frac{p^2}{ t_2} +\frac{p^4}{{t_2}^2} +\cdots\right) \right]\ .
\label{eq:2loop-scaled3}
\end{align}
Now combining the leading order terms in the expansions of Eqs.\eqref{eq:2loop-scaled1}, \eqref{eq:2loop-scaled2} and \eqref{eq:2loop-scaled3}, we find
\begin{equation}\label{eq:2loop-scaled4}
{I_2}_{\text{MS}}^{\text{leading}} =\frac{1}{m_u^2 m_{\tilde{R}_2}^2}[-J(r,t_1)+J(r,t_2)]
\end{equation}
where
\begin{equation}
J(r,t) =  \int d^4 p \int d^4 q \frac{1}{p^2q^2}  
\frac{1}{q^2-r}\frac{1}{(p-q)^2-1} \frac{1}{p^2-t}\ ,
\label{eq:2loop-int}
\end{equation}
is a standard two-loop integral, which can then be evaluated using dimensional regularisation, see for instance~\cite{vanderBij:1983bw,McDonald:2003zj,Angel:2013hla,AristizabalSierra:2014wal}. The closed form for ${I_2}_{\text{MS}}^{\text{leading}}$ is still fairly lengthy. However, under the approximation that logarithms of the masses of the SM fields and order unity factors can be ignored compared to the logarithms of heavy NP masses, we find the approximate form for ${I_2}_{\text{MS}}^{\text{leading}}$ given by
\begin{equation}\label{eq:2loop_SMS}
{I_2}_{\text{MS}}^{\text{leading}} \simeq  \frac{\pi^4}{m_{\tilde{R}_2}^2} \left[ 
%-\frac{1}{m_{\tilde{R}_2}^2}+\frac{1}{ m_{S_1}^2}
- \frac{c_n+\log\left(\frac{m_{\tilde{R}_2}^2}{m_W^2}\right)+\frac 12 \log^2 \left(\frac{m_{\tilde{R}_2}^2}{m_W^2}\right)}{m_{\tilde{R}_2}^2}+
\frac{ c_n+\log\left(\frac{m_{S_1}^2}{m_W^2}\right)+\frac 12 \log^2 \left(\frac{m_{S_1}^2}{m_W^2}\right)}{m_{S_1}^2}\right]\, ,
\end{equation}
where $c_n\equiv 1+\pi^2/3$ and we have taken the limit of vanishing $m_u$ after evaluating the integrations in Eq.~\eqref{eq:2loop-scaled4}. The approximation for the neutrino mass above matches very well with the full model exact loop computation of Eq. \eqref{eq:thefinalnumass}, in the limit of the hierarchical new physics scales being much heavier than the SM masses in the loops. Note that while the above approximation works well when $m_{\tilde{R}_2}\gg m_{S_1}$, it leads to a suppression with respect to the full model loop computation when $m_{\tilde{R}_2} \sim m_{S_1}$. This is due to the cancellation of the symmetric structure in the square bracket in Eq.~\eqref{eq:2loop_SMS}. This apparent mismatch can be mitigated by making Eq.~\eqref{eq:2loop_SMS} symmetric by replacing $m_{\tilde{R}_2}^2$ by $(m_{\tilde{R}_2}^2-m_{S_1}^2)$ in the prefactor, which is the form that we will use hereafter.

\subsection{Generalised simplified multi-scale results for radiative neutrino masses}
 Given the above discussion, showing the effectiveness of the dimensional regularisation based approach, we can now generalise our results for any tree-level realisation of the dimension-7 SMEFT operators leading to loop-level neutrino masses via diagrams shown in Fig.~\ref{fig:topsIandII_sim}. Depending on the actual realisation, there can be additional colour factors, symmetry factors, and even swapping (between mass and momentum exchange) of some mass scales, depending on the projection operators in the diagram.  To keep the results as general as possible, we will denote the heavy new bosons in the left diagram of Fig.~\ref{fig:topsIandII_sim} by $\Phi_1$ and $\Phi_2$ . For topology I, the relevant simplified interactions are then given by the Lagrangian
\begin{equation}
	\label{eq:lagSimpMI}
	\mathcal{L}^\text{I}=\mathcal{L}_\text{SM}+\mathcal{L}^\text{kin}_{\Phi_1}+\mathcal{L}^\text{kin}_{\Phi_2} -\lambda_{\Phi_1} f L \Phi_1^* -\lambda_{\Phi_2} f L \Phi_2^*-\mu\Phi_1^*\Phi_2^* H\, ,
\end{equation}
where $\mu$ is a dimensionful coupling, and $\mathcal{L}^\text{kin}_{\Phi_1}$ and $\mathcal{L}^\text{kin}_{\Phi_2}$ are the kinetic Lagrangians for $\Phi_1$ and $\Phi_2$, respectively. Similarly, for topology II, we consider a simplified model that contains one heavy new fermion $\Psi$ and one heavy new scalar $\Phi$ interacting with the SM Higgs and lepton doublets, respectively. This scenario can be described by the simplified Lagrangian
\begin{equation}
	\label{eq:lagSimpMII}
\mathcal{L}^\text{II}=\mathcal{L}_\text{SM}+\mathcal{L}^\text{kin}_\Psi+\mathcal{L}^\text{kin}_\Phi-\lambda_f f L \Phi^*-y_\Psi f \bar\Psi H\, ,
\end{equation}
where $\lambda_f$ and $y_\Psi$ are dimensionless coupling constants, and where $f$ is a SM fermion. 

The two new heavy BSM fields that are introduced in either case will be assumed to have a large hierarchy. By neglecting the masses of the SM fermions in the loop and any colour or symmetry factors, we can approximate the radiative neutrino mass generated in topology I by 
\begin{equation}
	\label{eq:numassUVI}
	(m_\nu)^\text{I}_{ij}\approx \frac{1}{16\pi^2}\frac{v\mu}{ m_{\Phi_1}^2-m_{\Phi_2}^2}\log\left(\frac{m_{\Phi_1}^2}{m_{\Phi_2}^2}\right)\left(\lambda_{\Phi_1} M_f\lambda_{\Phi_2}^T+\lambda_{\Phi_2}M_f^T\lambda_{\Phi_1}^T\right)_{ij}\, ,
\end{equation} 
 and topology II by
	\begin{equation}
		\label{eq:numassUVII}
		(m_\nu)^\text{II}_{ij}\approx \frac{1}{16\pi^2}\frac{v}{ m_{\Psi}^2-m_{\Phi}^2}\log\left(\frac{m_{\Psi}^2}{m_{\Phi}^2}\right)\left(\lambda_{f} M_fy_{\Psi}M_\Psi^T\lambda_{\Psi}^T+\lambda_{\Psi} M_\Psi y_{\Psi}^TM_f^T\lambda_{f}^T\right)_{ij}\, .
\end{equation}
Here $m_\alpha$ for $\alpha\in\{\Phi,\Psi,\Phi_1,\Phi_2\}$ is the mass of the corresponding heavy BSM field, and $M_f$ and $M_\Psi$ are the diagonal $3\times 3$ mass matrices for the SM fermion $f$ and BSM fermion $\Psi$ respectively, where we have assumed that $\Psi$ has three generations. At two-loop, we have
\begin{equation}
\begin{aligned}
	\label{eq:numassUVIloop}
	(m_\nu)^\text{I, two-loop}_{ij} &\approx \frac{1}{(16\pi^2)^2}\frac{v^3\mu}{m_{\Phi_1}^2-m_{\Phi_2}^2}\left[ f_{\ell}(m_{\Phi_1},m_W)-f_{\ell}(m_{\Phi_2},m_W)\right]\\
    &\times\left(\lambda_{\Phi_1} M_f\lambda_{\Phi_2}^T+\lambda_{\Phi_2}M_f^T\lambda_{\Phi_1}^T\right)_{ij}\, 
    \end{aligned}
\end{equation} 
in topology I and
	\begin{equation}
    \begin{aligned}
		\label{eq:numassUVIIloop}
		(m_\nu)^\text{II, two-loop}_{ij} &\approx \frac{1}{(16\pi^2)^2}\frac{v^3}{m_{\Psi}^2-m_{\Phi}^2}\left[ f_{\ell}(m_\Psi,m_W)-f_{\ell}(m_\Phi,m_W)\right]\\
&\times\left(\lambda_{f} M_fy_{\Psi}M_\Psi^T\lambda_{\Psi}^T+\lambda_{\Psi} M_\Psi y_{\Psi}^TM_f^T\lambda_{f}^T\right)_{ij}\, 
        \end{aligned}
	\end{equation}
in topology II, where the function $f_{\ell}(m_\chi,m_W)$ is defined as

\begin{equation}
    \begin{aligned}
		\label{eq:2loop_fn}
		f_{\ell}(m_\chi,m_W)\equiv  \frac{\left(1+\frac{\pi^2}{3}\right) +\log\left(\frac{m_{\chi}^2}{m_W^2}\right)+\frac 12 \log^2 \left(\frac{m_{\chi}^2}{m_W^2}\right)}{m_{\chi}^2}\, .
        \end{aligned}
	\end{equation}

Besides generating radiative neutrino masses, the simplified models discussed here will also lead to other phenomenological processes involving the $\Delta L = 2$ dimension-7 operators (cf.\ Ref.~\cite{Fridell:2023rtr}). The corresponding Wilson coefficients can be written as 
\begin{equation}
	\label{eq:EFTscaleI}
	C_{ij}^\text{I}=\frac{\mu \left(\lambda_{\Phi_1} \lambda_{\Phi_2}\right)_{ij}}{m_{\Phi_1}^2m_{\Phi_2}^2}
\end{equation}
for topology I, and
\begin{equation}
	\label{eq:EFTscaleII}
	C_{ij}^\text{II}=\frac{\left(\lambda_{\Psi} y_\Psi\lambda_{f}\right)_{ij}}{m_{\Phi}^2m_{\Psi}}
\end{equation}
for topology II. The relevant phenomenology is briefly discussed in Sec.~\ref{sec:combUVLNV}. 

The results for the new simplified multi-scale approach, as derived in the previous subsection and generalised in this section, applied to the model example discussed in Sec.~\ref{sec:leptoquark}, are shown in Fig.~\ref{fig:comparemnu} using red solid lines. Compared to the cut-off regularisation based estimates shown in green solid lines, it is easy to see that our simplified multi-scale prescription provides a much closer estimate of the exact model results. For one-loop radiative neutrino masses, the improvement is clearly visible when the ratio between the NP scales exceeds a few. On the other hand, for the two-loop radiative neutrino mass diagram, the results are even more striking. As discussed in Sec.~\ref{sec:doomed_cut-off}, the naive cut-off regularisation based EFT approach leads to an overestimation of the neutrino mass because of a cancellation in the leading contributions, even in the absence of hierarchy between the heavy new physics scales.

Generally, the simplified multi-scale approach based on dimensional regularisation reproduces the correct power counting behaviour as expected and leads to an excellent agreement with the full model exact computation for both hierarchical and non-hierarchical heavy new physics scenarios. While our dimensional regularisation based treatment is generalised to have two hierarchical heavy scales, the degenerate limit for the heavy scales of our general results, or evaluating the integrals directly assuming the two heavy scales to be the same, provides a consistent and relatively easy way to estimate the loop contributions consistently. An alternative consistent approach would be to use the complete basis of the EFT dimension by dimension, supplemented by the corresponding RGE equations and the explicit matching of the UV model at the scale of EFT.

 Therefore, we conclude that the treatment of radiative neutrino masses using the simplified multi-scale approach proposed in this work leads to a very accurate approximation of the exact results in spite of a number of approximations and simplifications with respect to the exact full model computation, thereby demonstrating the simplicity and power of the simplified multi-scale approach. In what follows, we now show that allowing for the masses of the heavy new physics fields, realising higher dimensional LNV operators to be  independent parameters, opens up new parts
of phenomenological parameter space, which can be explored and confronted with a number of LNV experimental observables.

%%%%%%%%%%%%%%%%%%%%%%%%%%%%%%%%%%%%%%%%%%%%%%%%%%%%%%%%%%%%%%%%%%%%%%%%%%%%%%%%%
%%%%%%%%%%%%%%%%%%%%%%%%%%%%%%%%%%%%%%%%%%%%%%%%%%%%%%%%%%%%%

%%%%%%%%%%%%%%%%%%%%%%%%%%%%%%%%%%%%%%%%%%
\section{Phenomenology of UV-completions at dimension-7}
\label{sec:combUVLNV}
%%%%%%%%%%%%%%%%%%%%%%%%%%%%%%%%%%%%%%%%%%%%%%%%%%%%%%%%%%%%%%%%%%%%%%%%%%%%%%%%%

\begin{figure}
	\centering
	\includegraphics[width=0.365\textwidth]{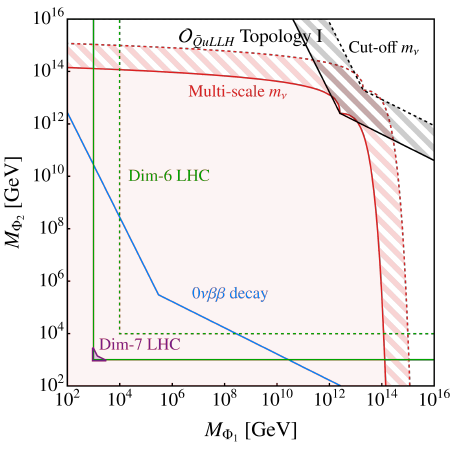}
	\includegraphics[width=0.365\textwidth]{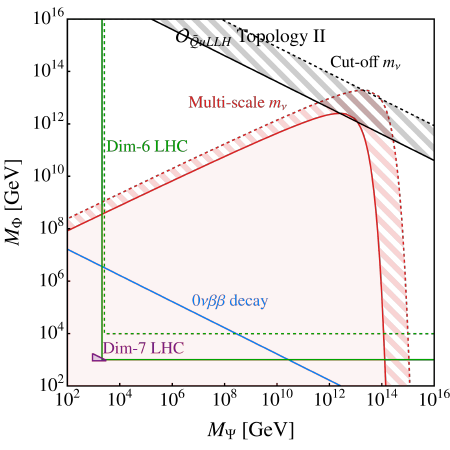}\\
\includegraphics[width=0.365\textwidth]{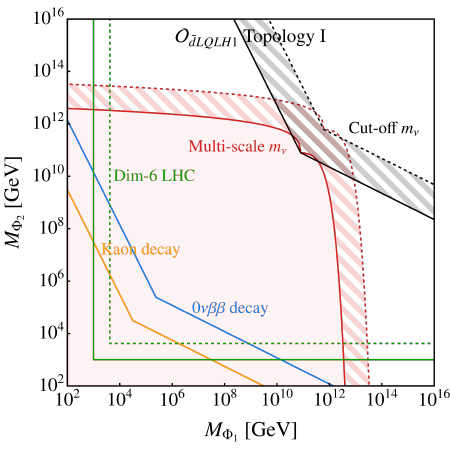}
\includegraphics[width=0.365\textwidth]{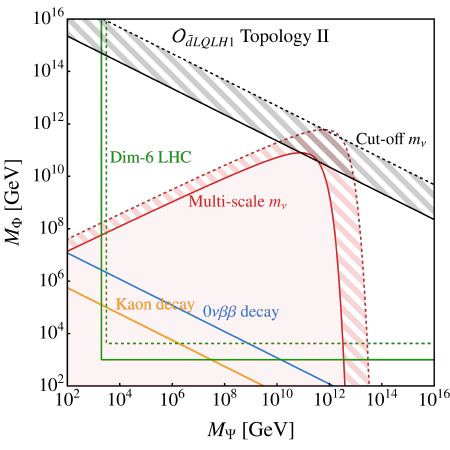}\\
\includegraphics[width=0.365\textwidth]{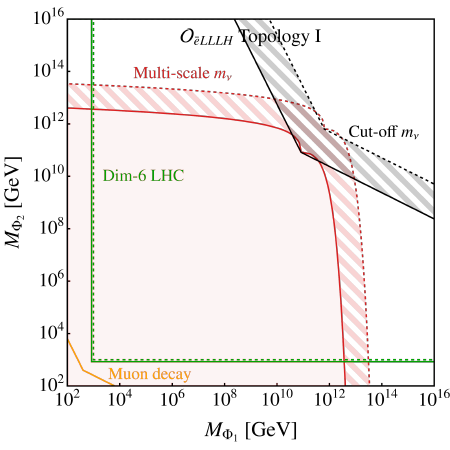}
\includegraphics[width=0.365\textwidth]{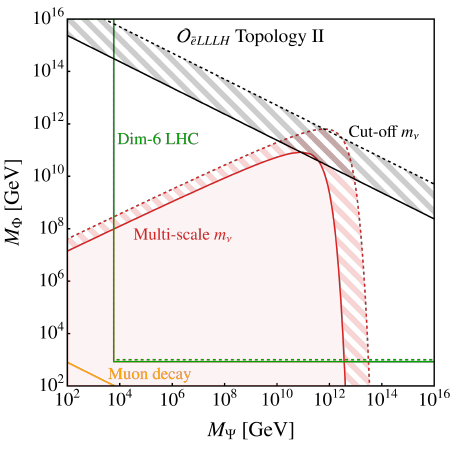}
	\caption{Parameter space for operator $\mathcal{O}_{\bar QuLLH}$ (top row), $\mathcal{O}_{\bar dLQLH1}$ (center row), and $\mathcal{O}_{\bar eLLLH}$ (bottom row) in topology~I (left column) and topology~II (right column). The red and black solid curves correspond to the exclusion limits from neutrino masses based on Eq.~\eqref{con:KATRIN} and oscillation data constraints for normal ordering using the multiscale approach proposed in this work (exclusion is indicated by the red shaded area), and the naive cut-off estimates in Tab.~\ref{tab:dim7opExp}, respectively. The corresponding dotted lines are obtained for $m_1=0$. The hashed regions are allowed areas. Further exclusion limits from various observables are shown with the area to the left and below excluded.}
	\label{fig:contour_QuLLH}
\end{figure}

\begin{figure}
	\centering
\includegraphics[width=0.396\textwidth]{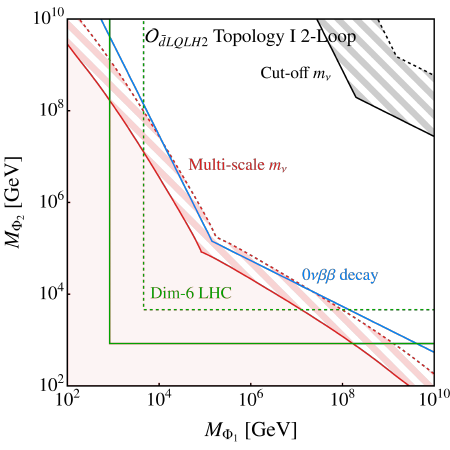}
\includegraphics[width=0.396\textwidth]{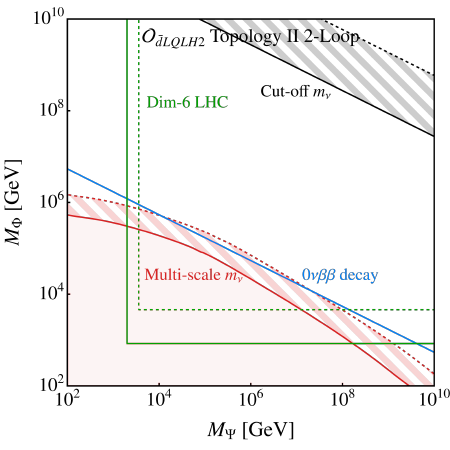}\\
	\includegraphics[width=0.396\textwidth]{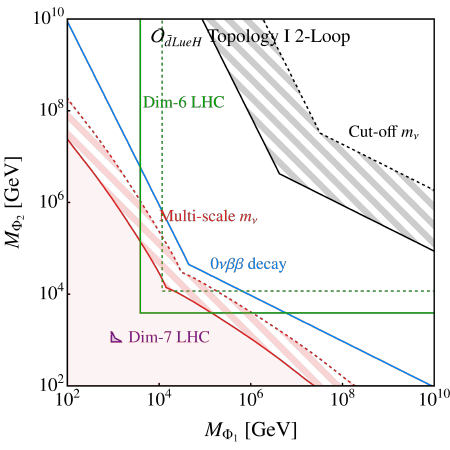}
	\includegraphics[width=0.396\textwidth]{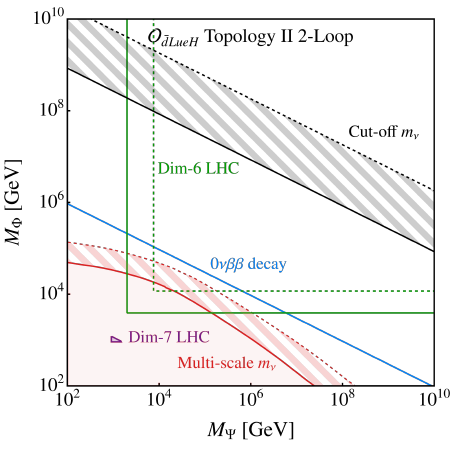}
	\caption{Same as Fig.~\ref{fig:contour_QuLLH}, but for operator $\mathcal{O}_{\bar dLQLH2}$ (top row) and $\mathcal{O}_{\bar dLueH}$ (bottom row).
 }
	\label{fig:contour_dLueH}
\end{figure}

In this section, we discuss the different constraints on the simplified models that appear as UV-completions of the $\Delta L=2$ dimension-7 operators, including collider constraints as well as low-energy observables such as neutrinoless double beta ($0\nu\beta\beta$) decay, and compare these constraints to radiative neutrino mass generation using the different approaches discussed above. In Figs.~\cref{fig:contour_QuLLH,fig:contour_dLueH} we show the constraints in the parameter space spanned by the two heavy scales, for tree-level UV-completions of the four-fermion $\Delta L=2$ dimension-7 effective operators for both topologies I and~II. We chose this plane since it best highlights the effects of a varying hierarchy in the internal degrees of freedom. Here, all dimensionless couplings are set to unity, and the dimensionful coupling $\mu$ is parametrised as $\mu=\text{max}(m_{\Phi_1}, m_{\Phi_2})$. We focus on operators $\mathcal{O}_{\bar QuLLH}$, $\mathcal{O}_{\bar dLueH}$, $\mathcal{O}_{\bar dLQLH1}$, $\mathcal{O}_{\bar dLQLH2}$, and $\mathcal{O}_{\bar eLLLH}$ in this section.

In Ref.~\cite{Fridell:2023rtr}, we presented the collider limits for dimension-7 operators when the relevant interactions are taken as point-like, i.e.\ when the heavy NP degrees of freedom are not produced on-shell as resonances. 
The purple regions in Figs.~\cref{fig:contour_QuLLH,fig:contour_dLueH}, corresponding to constraints from searches for $pp\to\ell^+\ell^+jj$ at the LHC, are cut off at the lower ends in both NP mass-variables due to the constraint $M<900$~GeV as required to ensure the validity of the EFT-approach (c.f.\ Ref.~\cite{Fridell:2023rtr}). The resulting constrained areas are comparatively small and are always fully excluded by the flavour-universal neutrino mass constraint. For operators $\mathcal{O}_{\bar dLQLH1}$ and $\mathcal{O}_{\bar dLQLH2}$ the area vanishes completely, as seen in Figs.~\ref{fig:contour_QuLLH} (center row) and Fig.~\ref{fig:contour_dLueH} (top row), respectively, since the constraint is lower than the EFT validity scale for all NP masses. However, the cross section for $pp\to\ell^+\ell^+jj$ may be enhanced by on-shell mediators, such as e.g.\ in the Keung-Senjanović process~\cite{Keung:1983uu}.

Under the assumption that the $\Delta L = 2$ dimension-7 operators are generated at tree-level, collider constraints on the individual heavy fields that generate a given operator will translate into constraints on the Wilson coefficient of the operator itself. Collider searches for new heavy fields are typically most sensitive when the field can be created on-shell as a resonance. However, such resonant searches are typically dependent on model parameters. Instead, constraints on the underlying new heavy particles in the different UV-completions can also be derived from global fits of dimension-6 operators~\cite{Ellis:2020unq,Crivellin:2021bkd}. The constraints obtained from this method are typically within an order of magnitude of the constraints from resonant searches. We use the 95\% C.L.\ constraints on the masses of the different new fields that appear in the UV-completions of $\Delta L = 2$ dimension-7 operators from Refs.~\cite{Ellis:2018gqa,Ellis:2020unq,Crivellin:2021bkd}. For these constraints, we assume flavour universality and take the most stringent constraints for the cases where several results are available in the literature. 
In the following analysis, in order to reduce the number of parameters, we set all couplings to unity. The green lines shown in Figs.~\cref{fig:contour_QuLLH,fig:contour_dLueH} indicate the region excluded by global fits of dimension-6 operators for LHC searches, translated into limits on the corresponding tree-level UV-completions~\cite{Ellis:2020unq,Crivellin:2021bkd}. The dashed green lines denote the most stringent constraint, and the solid green lines the least stringent.
\begin{table}[]
    \centering
    \setlength{\extrarowheight}{8pt}
    \begin{tabular}{| l l |}
    \hline
    \specialrule{2pt}{2pt}{0pt}
         \multicolumn{2}{| c |}{\cellcolor{newgray2} \raisebox{2pt}{Cuts for $p p \to \mu^\pm\mu^\pm j j$ at $\sqrt{s}=13$ TeV} } \\
         \hline
         Object selection cuts & \\
         \hline
         $p_T^{\mu^{1(2)}}>25$ GeV & $p_T^{j^{1(2)}}>20$ GeV \\
         $|\eta^{\mu^{1(2)}}|<2.5$ & $|\eta^{j^{1(2)}}|<2.5$ \\
         \hline
         Track-to-vertex association cuts & \\
           \hline
         $|z_0 \sin \theta|< 5$ mm & $|d_0|< 1$ {\textmu}m\\
         \hline
         Signal region cuts & \\
         \hline
         $p_T^{\mu^{\text{leading}}}>40$ GeV & $p_T^{j^{1(2)}}>100$ GeV \\
         $H_T> 400$ GeV  & $\Delta R_{\mu\mu}< 3.9$ GeV \\
          \raisebox{4pt}{$m_{\mu^{1}\mu^{2}}>400$ GeV} & \raisebox{4pt}{$m_{j^{1}j^{2}}>110$ GeV}\\
    \specialrule{2pt}{0pt}{2pt}
           \hline
    \end{tabular}
    \caption{Cuts used to calculate the number of $pp\to\ell^\pm\ell^\pm$ events at
    ATLAS in the $\tilde R_2$-$S_1$ leptoquark model, following Ref.~\cite{ATLAS:2023cjo}.}
    \label{tab:cuts}
\end{table}

The model example discussed in Sec.~\ref{sec:leptoquark} is subject to constraints from leptoquark searches at the LHC. The leptoquarks $S_1$ and $\tilde R_2$ both lead to distinct signatures at the LHC, where they can be produced e.g.\ via pair production $pp\to \text{LQ }\text{LQ}$ or single production $pp\to\ell\text{ LQ}$, see e.g.~Refs.~\cite{Blumlein:1996qp,Schmaltz:2018nls}. Using these modes, the masses of $\tilde R_2$ and $S_1$ have been constrained to $m_{\tilde R_2}<3.4$~TeV and $m_{S_1}<4.6$~TeV, respectively, at $95$\% C.L.~\cite{Bhaskar:2023ftn}. However, the above-mentioned processes are not LNV by two units $\Delta L = 2$, and therefore they do not directly correspond to Majorana neutrino masses. In order to probe the latter scenario, an overall $\Delta L=2$ process such as $pp\to\ell^\pm\ell^\pm jj$ would need to be searched for. Such an analysis was performed for the dimension-7 LNV operators in Ref.~\cite{Fridell:2023rtr} from a model-independent perspective, leading to constraints $\Lambda_\text{LNV} > 1.1$~TeV on both of the operators that are generated in the $\tilde R_2$-$S_1$ model, where $\Lambda_\text{LNV}$ is the corresponding NP scale. However, these operator-based constraints could be an underestimate of what the actual model-dependent constraints would be, since a given model can lead to resonant enhancements. To find the model-dependent constraints on the $\tilde R_2$-$S_1$ model coming from $pp\to\ell^\pm\ell^\pm jj$ searches at the LHC, we implement the model using \textsc{FeynRules}~\cite{Alloul:2013bka} and calculate the cross section with \textsc{MadGraph5}~\cite{Alwall:2014hca}, including \textsc{Pythia8}~\cite{Sjostrand:2014zea} for hadronisation and using the PDF set \textsc{NNPDF30} provided by \textsc{LHAPDF6}~\cite{Buckley:2014ana}, after which we perform detector simulation using \textsc{Delphes3}~\cite{deFavereau:2013fsa} with the cuts shown in Tab.~\ref{tab:cuts}, and compare with the ATLAS results~\cite{ATLAS:2023cjo} for LHC Run 2 at $\sqrt{s}=13$~TeV. We perform this analysis for $\mu=\max (m_{\tilde R_2},m_{S_1})$ in two scenarios: a) $g_1=g_2=1$ and $g_3=0$, corresponding to $\mathcal{O}_{\bar dLQLH1}$, and b) $g_1=g_3=1$ and $g_2=0$, corresponding to $\mathcal{O}_{\bar dLueH}$. For a vanishing hierarchy, such that $m_{\tilde R_2}=m_{S_1}\equiv \Lambda_\text{LNV}$, and at 95\% C.L., we find $\Lambda_\text{LNV}>3.1$~TeV in both scenarios a) and b).

There is a variety of ways that LNV can be probed at low energies, as was discussed in detail in Ref.~\cite{Fridell:2023rtr}. The most stringent bounds generally stem from neutrinoless double beta (\ovbb$\!\!\!$) decay searches, which, however, can probe only the first-generation couplings. Therefore, in NP models that dominantly couple the second or third-generation fermions, \ovbb decay searches might still not lead to any signals. In such models, it may be rare kaon decays that provide the first hints of LNV~\cite{Li:2019fhz,Deppisch:2020oyx,Buras:2024ewl}. Nonetheless, at leading order and within dimension-7 SMEFT, these processes constrain only operator $\mathcal{O}_{\bar dLQLH1}$. Further, operator $\mathcal{O}_{\bar eLLLH}$ does not lead to \ovbb decay (at tree-level) and may instead be probed in lepton-number-violating rare muon decays~\cite{Cirigliano:2017djv}. Other observables are not discussed in this work due to them being generally sub-dominant, e.g.\ coherent elastic neutrino-nucleus scattering (CE$\nu$NS), long-baseline neutrino oscillation, $\mu^-$ to $e^+$ conversion, and other LNV meson decays such as $K^+\to\pi^-\ell^+\ell^+$ or $B\to K\nu\nu$ (see Ref.~\cite{Fridell:2023rtr}). The blue line shows the area excluded by \ovbb decay, and orange line shows the exclusion from rare kaon decays for $\mathcal{O}_{\bar dLQLH1}$ (Fig.~\ref{fig:contour_QuLLH} center row), and the same colored line shows the exclusion from rare muon decays for $\mathcal{O}_{\bar eLLLH}$ (Fig.~\ref{fig:contour_QuLLH} bottom row).

The red regions in Figs.~\cref{fig:contour_QuLLH,fig:contour_dLueH} show the area of parameter space where the sum of neutrino masses is larger than the limit obtained from the KATRIN experiment~\cite{KATRIN:2001ttj}\footnote{Note that we do not use the corresponding PLANCK limits~\cite{Planck:2018vyg} due to their dependence on the underlying cosmological model.} following the simplified multi-scale approach described in Sec.~\ref{sec:dim_reg}. Note that the constraint from KATRIN is set on $m_\nu^\text{KATRIN}$ where $(m_\nu^\text{KATRIN})^2=\sum_i|U_{e i}|^2m_i^2<(0.45$~eV$)^2$~\cite{Katrin:2024tvg}, where $U$ is the PMNS matrix. We translate this into a limit on the sum of neutrino masses Tr $\hat m_\nu $ by using the central values of the neutrino mixing angles as well as the mass splittings $\Delta m_{12}^2 = 7.49\times 10^{-5}$~eV$^2$ and $\Delta m_{31}^2 = 2.513\times 10^{-3}$~eV$^2$ given by \textsc{NuFIT v6.0}~\cite{Esteban:2024eli}, assuming normal ordering, and solving for the free parameter $m_1$ (the lightest neutrino mass), such that
\begin{equation}{\label{con:KATRIN}}
    |U_{e1}|^2m_1^2+|U_{e2}|^2(m_1^2+\Delta m_{12}^2)+|U_{e3}|^2(m_1^2+\Delta m_{13}^2)<(m_\nu^\text{KATRIN})^2\, .
\end{equation}
A lower limit can be obtained by setting the lowest neutrino mass $m_1$ to zero. The area between these two limits is shown as a striped red region in Figs.~\cref{fig:contour_QuLLH,fig:contour_dLueH}. The black striped region shows the corresponding neutrino mass constraints using the conventional cut-off approximation. Note that the dip in the top-left part of the neutrino mass constraint for topology~I comes from our choice of $\mu=\max (m_{\Phi_1},m_{\Phi_2})$.

\begin{figure}
	\centering
\includegraphics[width=0.396\textwidth]{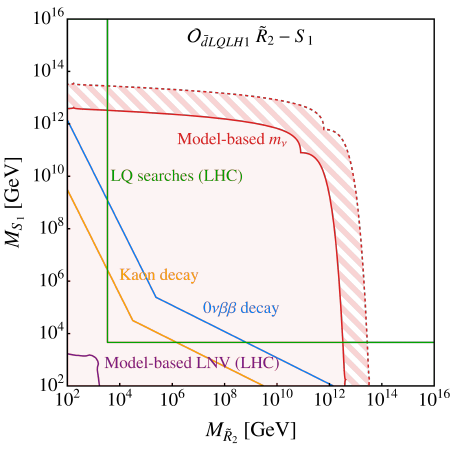}
\includegraphics[width=0.396\textwidth]{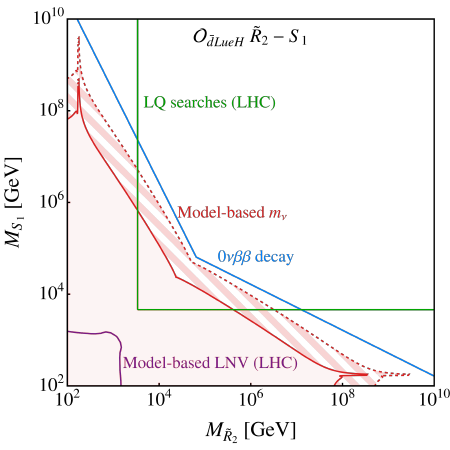}
	\caption{Same as Fig.~\ref{fig:contour_QuLLH}, but for the $\tilde R_2 - S_1$ model described in Sec.~\ref{sec:leptoquark}.}
	\label{fig:contour_R2tS1}
\end{figure}

In Figs.~\cref{fig:contour_QuLLH,fig:contour_dLueH} we see that the constraint on 1-loop neutrino masses using the cut-off estimate (black striped area) generally only agrees with our simplified model approach (red striped area) for a vanishing hierarchy, i.e.\ at the point where the two masses are approximately equal (along the diagonal from bottom left to top right). For a larger hierarchy, the disagreement increases, which is also what was found in the comparison in Fig.~\ref{fig:comparemnu}. Fig.~\ref{fig:contour_dLueH} corresponds to operators for which the neutrino mass is generated at 2-loop order. Here we see that the cut-off estimate does not reproduce the results of our more accurate simplified-model approach for most of the parameter space, as seen in Fig.~\ref{fig:comparemnu}. One reason for this mismatch is that the 2-loop mass generally depends on all mass scales involved, rather than just the highest scale, as was our approximation in the 1-loop case. We further see that the regions which lead to the observed neutrino mass are generally excluded by \ovbb decay.

For some operators, the single-particle limits from global fits of LHC results (green lines) roughly overlap with the \ovbb and neutrino mass constraints simultaneously, in regions of parameter space where one field is significantly more massive than the other, i.e.\ where there is a large hierarchy. 

In Fig.~\ref{fig:contour_R2tS1} we show the different constraints in the leptoquark model presented in Sec.~\ref{sec:leptoquark}, for the realisation of operators $\mathcal{O}_{\bar dLQLH1}$ (left) and $\mathcal{O}_{\bar dLueH}$ (right). Note that for the latter case, the neutrino mass is generated at 2-loop order, while at 1-loop order as in the former. The different colored lines (areas) correspond to experimental constraints similar to Figs.~\ref{fig:contour_QuLLH} and~\ref{fig:contour_dLueH}, i.e.\ green, purple, orange, blue and red correspond to direct LHC searches for $\tilde R_2$ and $S_1$, LNV searches at the LHC, rare kaon decay, \ovbb decay,  and neutrino masses, respectively. In contrast, the striped red areas correspond to the observed neutrino masses. The two different operators are generated by two different couplings of $S_1$ to the SM. In both cases, the limits coming from LNV searches at the LHC (purple) are subdominant compared to direct searches (green) for the whole parameter space. For comparisons with our simplified model approach, the corresponding figures are Fig.~\ref{fig:contour_QuLLH} (centre left) for Fig.~\ref{fig:contour_R2tS1} (left) and Fig.~\ref{fig:contour_dLueH} (bottom left) for Fig.~\ref{fig:contour_R2tS1} (right).

The accuracy of the simplified multi-scale approach introduced in Sec.~\ref{sec:dim_reg} can be, again, verified by its comparison with the UV-complete scenario. Specifically, the neutrino-mass constraints shown in red in the left-centre panel of Fig.~\ref{fig:contour_QuLLH} and in the left-bottom panel of Fig.~\ref{fig:contour_dLueH} are in excellent agreement with the bounds obtained from the full model depicted in Fig.~\ref{fig:contour_R2tS1}, left and right panels, respectively. We also see in Figs.~\Cref{fig:contour_QuLLH,fig:contour_dLueH} that a hierarchy in the internal degrees of freedom of a given UV-completion of a dimension-7 $\Delta L=2$ operator has a significant effect on the size of the available parameter space. Unlike the naive single-scale cut-off-regularisation-based limits, the more realistic neutrino mass bounds obtained by the simplified multi-scale approach open up viable regions of parameter space within potential reach of LHC searches and/or future \ovbb decay experiments.

%%%%%%%%%%%%%%%%%%%%%%%%%%%%%%%%%%%%%%%%%%%%%%%%%%%%%%%%%%%%%%%%%%%%%%%%%%%%%%%%%
\section{Conclusions}
\label{sec:conclusions}
%%%%%%%%%%%%%%%%%%%%%%%%%%%%%%%%%%%%%%%%%%%%%%%%%%%%%%%%%%%%%%%%%%%%%%%%%%%%%%%%%
In this work, we have extended the phenomenological investigations of dimension-7 $\Delta L = 2$ operators in the framework of SMEFT by analysing their UV-completions, in particular, concerning the generation of the observed neutrino masses. Generally, dimension-7 lepton-number-violating interactions are, following the usual EFT paradigm, expected to be more suppressed than the well-known dimension-5 Weinberg operator. However, in UV scenarios without the seesaw fields, these operators can provide the leading contribution to LNV processes and provide a valuable insight into radiative Majorana neutrino mass models.

After obtaining the full list of tree-level UV-completions of dimension-7 $\Delta L=2$ operators, we identified the radiative neutrino Majorana mass topologies corresponding to these UV-completions, and provided approximate expressions for the value of the neutrino mass in both scenarios. When doing so, we have used a dimensional-regularisation-based approach to study neutrino mass contributions in a multiscale setup, ensuring a more robust analysis and allowing for a realistic description in the presence of a hierarchy among new fields. We then compared our results to a leptoquark model example, for which we find very good agreement and a significant improvement over conventional approximations of the neutrino mass corresponding to dimension-7 operators employing naive cut-off-based regularisation. Most notably, we showed that a hierarchy in the internal degrees of freedom of an operator will significantly affect the neutrino mass. 

As part of the subsequent phenomenological analysis, we compared our results to other observables probing LNV, including \ovbb decay and the LHC. Among other things, we found that there are regions of parameter space for most operators that successfully generate the observed neutrino mass while being close to the experimental reach of both the LHC and \ovbb decay. These regions generally involve a large internal hierarchy, and as such have evaded attention in previous studies, assuming a common scale of NP. In this way, our findings provide valuable guidance for future experimental efforts aimed at probing LNV and searching for BSM physics.

%%%%%%%%%%%%%%%%%%%%%%%%%%%%%%%%%%%%%%%%%%%%%%%%%%%%%%%%%%%%%%%%%%%%%%%%%%%%%%%%%
\section*{Acknowledgements}
%%%%%%%%%%%%%%%%%%%%%%%%%%%%%%%%%%%%%%%%%%%%%%%%%%%%%%%%%%%%%%%%%%%%%%%%%%%%%%%%%
K.~F. acknowledges support from the Japan Society for the Promotion of Science (JSPS)
Grant-in-Aid for Scientific Research B (No. 21H01086).
L.~G. acknowledges support from the Dutch Research Council (NWO), under project number VI.Veni.222.318; from Charles University through project PRIMUS/24/SCI/013; from the National Science Foundation, Grant PHY-1630782, and the Heising-Simons Foundation, Grant 2017-228. L.~G. is also thankful to the Instituto de Física Corpuscular (IFIC) for the hospitality provided during his visit that led to the finalisation of this project. 
J.~H.\ acknowledges support by the Cluster of Excellence “Precision Physics, Fundamental Interactions, and
Structure of Matter” (PRISMA$^+$ EXC 2118/1) funded by the Deutsche Forschungsgemeinschaft (DFG, German Research
Foundation) within the German Excellence Strategy (Project No. 390831469). 
C.~H. is funded by the Generalitat Valenciana under Plan Gen-T via CDEIGENT grant No. CIDEIG/2022/16.  C.~H.\ also acknowledges support from the Spanish grants PID2023-147306NB-I00 and CEX2023-001292-S (MCIU/AEI/10.13039/501100011033).
K.~F., J.~H. and C.~H. also acknowledge support from the Emmy Noether grant "Baryogenesis, Dark Matter and Neutrinos: Comprehensive analyses and accurate methods in particle cosmology" (HA 8555/1-1, Project No. 400234416) funded by the Deutsche Forschungsgemeinschaft (DFG, German Research Foundation).

%%%%%%%%%%%%%%%%%%%%%%%%%%%%%%%%%%%%%%%%%%%%%%%%%%%%%%%%%%%%%%%%%%%%%%%%%%%%%%%%%
%%%%%%%%%%%%%%%%%%%%%%%%%%%%%%%%%%%%%%%%%%%%%%%%%%%%%%%%%%%%%%%%%%%%%%%%%%%%%%%%%
\appendix
\section{Covariant derivative expansion}\label{sec:CDE}
In order to relate the $\Delta L=2$ operators to the underlying tree-level UV-completions we here employ the covariant derivative expansion formalism (CDE)~\cite{Henning:2014wua} following Ref.~\cite{Gargalionis:2020xvt}. In this Appendix we show the calculations corresponding to the results presented in Sec.~\ref{sec:eft}, for a scalar, fermion, and vector field, respectively.

\subsection{Scalar field}\label{sec:cdes}
A general Lagrangian that contain both light fields $\pi_i$ and heavy complex scalar fields $\Phi$ can be written as
\begin{equation}
	\label{eq:SLag}
	\mathcal{L}_\text{S} \supset \mathcal{L}_{\Phi}^{\text{int}}+\mathcal{L}_{\Phi}^{\text{kin}}=\left(\Phi\frac{\partial \mathcal{L}_\Phi^{\text{int}}}{\partial \Phi}+\text{h.c.}\right)+\Phi^*\left(-D^2-m_\Phi^2\right)\Phi + \mathcal{O}(\Phi^3)\, .
\end{equation}
Here $D$ is a covariant derivative, and we have neglected terms quadratic in $\Phi$. Evaluating the linearised equations of motion (EOM) for $\Phi$ then leads to
\begin{equation}
	\left(-D^2-m_\Phi^2\right)\Phi=-\frac{\partial \mathcal{L}_\Phi^{\text{int}}}{\partial \Phi^*} + \mathcal{O}(\Phi^2)\, ,
\end{equation}
which we can solve using the classical field
\begin{equation}
	\Phi_\text{cl}=\frac{1}{D^2+m_\Phi^2}\frac{\partial \mathcal{L}_\Phi^{\text{int}}}{\partial \Phi^*}\, .
\end{equation}
Expanding in $D^2/m_\Phi^2$ we have
\begin{equation}
	\begin{aligned}
		\Phi_\text{cl}&=\frac{1}{m_\Phi^2}\left(1+\frac{D^2}{m_\pi^2}\right)^{-1}\frac{\partial \mathcal{L}_\Phi^{\text{int}}}{\partial \Phi^*}=\left(\frac{1}{m_\Phi^2}-\frac{D^2}{m_\Phi^4}+\dots\right)\frac{\partial \mathcal{L}_\Phi^{\text{int}}}{\partial \Phi^*}\, .
	\end{aligned}
\end{equation}
Substituting the classical field into Eq.~\eqref{eq:SLag} gives us a series of interaction terms
\begin{equation}
	\label{eq:LeffScalarI}
	\mathcal{L}_{\text{eff}}\supset \mathcal{L}_{\text{eff}}^\Phi = \frac{\partial \mathcal{L}_\Phi^{\text{int}}}{\partial \Phi}\lr{\frac{1}{m_\Phi^2}-\frac{D^2}{m_\Phi^4} + \dots}\frac{\partial \mathcal{L}_\Phi^{\text{int}}}{\partial \Phi^*}\, .
\end{equation}
This expression contains an effective Lagrangian that is independent of the heavy scalar field $\Phi$, but that does depend on the interactions of $\Phi$ with other fields, both light and heavy, corresponding to the derivatives $\partial \mathcal{L}_\Phi^{\text{int}}/\partial \Phi^{(*)}$.

\subsection{Fermion field}\label{sec:cdef}
For a heavy Dirac fermion field $\Psi\equiv\left(\,\chi_\alpha\, , \, \eta^{\dagger\dot{\alpha}}\,\right)^T$ we have
\begin{equation}
	\begin{aligned}
		\label{eq:firstspinorlag}
		\mathcal{L}_\text{S}&\supset i\bar{\Psi}\slashed D\Psi-m_\Psi\bar{\Psi}\Psi +\left(\Psi\frac{\partial \mathcal{L}_\Psi^{\text{int}}}{\partial \Psi}+\text{h.c.}\right)\\
		&=i\chi^\dagger_{\dot{\alpha}} D^{\dot{\alpha}\beta} \chi_\beta+i\eta^\alpha D_{\alpha\dot{\beta}} \eta^{\dagger\dot{\beta}}+\left(\chi_\alpha\frac{\partial \mathcal{L}_\Psi^{\text{int}}}{\partial \chi_\alpha}+\eta^{\alpha}\frac{\partial \mathcal{L}_\Psi^{\text{int}}}{\partial \eta^{\alpha}}-m_\Psi\eta^\alpha\chi_\alpha+\text{h.c.}\right)\, .
	\end{aligned}
\end{equation}
Here we have used
\begin{equation}
	D^{\dot{\alpha}\beta}\equiv D^\mu\bar{\sigma}_{\mu}^{\dot{\alpha}\beta},\qquad D_{\alpha\dot{\beta}}\equiv D^\mu\sigma_{\mu\alpha\dot{\beta}}\, .
\end{equation}
The second term in Eq.~\eqref{eq:firstspinorlag} can be written in the form of the first using integration by parts,
\begin{equation}
	i\eta^\alpha D_{\alpha\dot{\beta}} \eta^{\dagger\dot{\beta}}=-i( D^\mu \eta^\alpha)\sigma_{\mu\alpha\dot{\beta}}\eta^{\dagger\dot{\beta}}\, .
\end{equation}
After applying a Fierz transformation
\begin{equation}
	-i( D^\mu \eta^\alpha)\sigma_{\mu\alpha\dot{\beta}}\eta^{\dagger\dot{\beta}}=i\eta^{\dagger}_{\dot{\beta}}\bar{\sigma}_{\mu}^{\dot{\beta}\alpha} D^\mu \eta_\alpha=i\eta^\dagger_{\dot{\beta}} D^{\dot{\beta}\alpha} \eta_\alpha\, ,
\end{equation}
we can write the EOM as
\begin{align}
	&i D^{\dot{\alpha}\beta} \chi_\beta-m_\Psi\eta^{\dagger\dot{\alpha}}+\frac{\partial \mathcal{L}_\Psi^{\text{int}}}{\partial \chi^\dagger_{\dot{\alpha}}}=0\label{eq:diraceom1}\, ,\\
	&i D^{\dot{\alpha}\beta} \eta_\beta-m_\Psi\chi^{\dagger\dot{\alpha}}+\frac{\partial \mathcal{L}_\Psi^{\text{int}}}{\partial \eta^\dagger_{\dot{\alpha}}}=0\label{eq:diraceom2}\, ,
\end{align}
where in Eq.~\eqref{eq:diraceom1} we can be solve for $\eta_{\dot{\beta}}^\dagger$,
\begin{equation}
	\begin{aligned}
		\eta_{\dot{\beta}}^\dagger=\frac{1}{m_\Psi}\epsilon_{\dot{\beta}\dot{\alpha}}\left(iD^{\dot{\alpha}\beta}\chi_\beta+\frac{\partial\mathcal{L}_\Psi^\text{int}}{\partial \chi_{\dot{\alpha}}^\dagger}\right)=i\frac{1}{m_{\Psi}}\epsilon_{\dot{\beta}\dot{\alpha}}D^{\dot{\alpha}\beta}\chi_{\beta}+\frac{1}{m_{\Psi}}\frac{\partial\mathcal{L}^\text{int}_\Psi}{\partial \chi^{\dagger\dot{\beta}}}\, .
	\end{aligned}
\end{equation}
Taking the complex conjugate of Eq.~\eqref{eq:diraceom2} now leads to
\begin{equation}
	\frac{1}{m_\Psi}D^{\dot{\beta}\alpha}\epsilon_{\dot{\beta}\dot{\alpha}}D^{\dot{\alpha}\beta}\chi_{\beta}+i\frac{1}{m_\Psi}D^{\dot{\beta}\alpha}\frac{\partial \mathcal{L}_\Psi^\text{int}}{\partial \chi^{\dagger\dot{\beta}}}-m_\Psi\chi^{\alpha}+\frac{\partial\mathcal{L}_\Psi^\text{int}}{\partial\eta_\alpha}=0\, .
	\label{eq:cdeferm1}
\end{equation}
Expressing the first term in Eq.~\eqref{eq:cdeferm1} using the field strength
\begin{equation}
	X_{\alpha}{}^{\beta}\equiv X^{\mu\nu}\sigma_{\nu\alpha\dot{\gamma}}\bar{\sigma}_{\mu}^{\dot{\gamma}\beta}=-i\left[D^\mu,\,D^\nu\right]\sigma_{\nu\alpha\dot{\gamma}}\bar{\sigma}_{\mu}^{\dot{\gamma}\beta}
\end{equation}
and the relation
\begin{equation}
	\begin{aligned}
		D^{\dot{\beta}\alpha}\epsilon_{\dot{\beta}\dot{\alpha}}D^{\dot{\alpha}\beta}&=D^\mu D^\nu \bar{\sigma}_\mu^{\dot{\alpha}\beta}\epsilon_{\dot{\beta}\dot{\alpha}}\bar{\sigma}_\nu^{\dot{\beta}\alpha}\\
		&=D^\mu D^\nu \epsilon^{\beta\gamma}\sigma_{\mu\gamma\dot{\beta}}\bar{\sigma}_\nu^{\dot{\beta}\alpha}\\
		&=D^\mu D^\nu\epsilon^{\beta\gamma}\left(\eta_{\mu\nu}\delta_{\gamma}{}^{\alpha}-2i(\sigma_{\mu\nu})_{\gamma}{}^{\alpha}\right)\\
		&=D^2\epsilon^{\beta\alpha}+\frac{i}{2}\epsilon^{\beta\gamma}X_\gamma{}^{\alpha}\, ,
	\end{aligned}
\end{equation}
where
\begin{equation}
	(\sigma_{\mu\nu})_\alpha{}^{\beta}\equiv
	\frac{i}{4} \left(\sigma_{\mu\alpha\dot{\gamma}}
	\bar{\sigma}_{\nu}^{\dot{\gamma}\beta}-\sigma_{\nu\alpha\dot{\gamma}}
	\bar{\sigma}_{\mu}^{\;\dot{\gamma}\beta}\right)\,,\quad
	(\bar{\sigma}_{\mu\nu})^{\dot{\alpha}}{}_{\dot{\beta}}\equiv
	\frac{i}{4}\left(\bar{\sigma}_\mu^{\dot{\alpha}\gamma}
	\sigma_{\nu\gamma\dot{\beta}}-\bar{\sigma}_\nu^{\dot{\alpha}\gamma}
	\sigma_{\mu\gamma\dot{\beta}}\right)\, .
\end{equation}
then leads to
\begin{equation}
	m_\Psi^2\left(\frac{-D^2\epsilon^{\alpha\delta}-\tfrac{i}{2}X_\gamma{}^{\alpha}\epsilon^{\gamma\delta}}{m_\Psi^2}-\epsilon^{\alpha\delta}\right)\chi_{\delta}+iD^{\dot{\beta}\alpha}\frac{\partial \mathcal{L}_\Psi^\text{int}}{\partial \chi^{\dagger\dot{\beta}}}+m_\Psi\frac{\partial\mathcal{L}_\Psi^\text{int}}{\partial\eta_\alpha}=0
\end{equation}
which we can solve using the classical fermion field
\begin{equation}
	\begin{aligned}
		(\chi_{\text{cl}})_\delta&=\frac{1}{m_\Psi^2}\left(\frac{-D^2\epsilon^{\alpha\delta}-\tfrac{i}{2}X_\gamma{}^{\alpha}\epsilon^{\gamma\delta}}{m_\Psi^2}-\epsilon^{\alpha\delta}\right)^{-1}\left(iD^{\dot{\beta}\alpha}\frac{\partial \mathcal{L}_\Psi^\text{int}}{\partial \chi^{\dagger\dot{\beta}}}+m_\Psi\frac{\partial\mathcal{L}_\Psi^\text{int}}{\partial\eta_\alpha}\right)\\
		&=\left(\frac{\epsilon_{\alpha\delta}}{m_\Psi^2}+\frac{D^2\epsilon_{\alpha\delta}+\tfrac{i}{2}X_\alpha{}^{\gamma}\epsilon_{\gamma\delta}}{m_\Psi^4}+\dots\right)\left(iD^{\dot{\beta}\alpha}\frac{\partial \mathcal{L}_\Psi^\text{int}}{\partial \chi^{\dagger\dot{\beta}}}+m_\Psi\frac{\partial\mathcal{L}_\Psi^\text{int}}{\partial\eta_\alpha}\right)\, .
	\end{aligned}
\end{equation}
The classical field $(\eta_{\text{cl}})_{\delta}$ can be obtained in an analogous way. We now substitute the fields in Eq.~\eqref{eq:firstspinorlag} to obtain a series of interaction terms
\begin{equation}
	\begin{aligned}
		\mathcal{L}_{\text{eff}}\supset\mathcal{L}_{\text{eff}}^\Psi &= \frac{\partial \mathcal{L}_\Psi^{\text{int}}}{\partial \chi_\delta}\left(\frac{\epsilon_{\alpha\delta}}{m_\Psi^2}+\frac{D^2\epsilon_{\alpha\delta}+\tfrac{i}{2}X_\alpha{}^{\gamma}\epsilon_{\gamma\delta}}{m_\Psi^4}+\dots\right)\left(iD^{\dot{\beta}\alpha}\frac{\partial \mathcal{L}_\Psi^\text{int}}{\partial \chi^{\dagger\dot{\beta}}}+m_\Psi\frac{\partial\mathcal{L}_\Psi^\text{int}}{\partial\eta_\alpha}\right)\\
		&+\frac{\partial \mathcal{L}_\Psi^{\text{int}}}{\partial \eta_\delta}\left(\frac{\epsilon_{\alpha\delta}}{m_\Psi^2}+\frac{D^2\epsilon_{\alpha\delta}+\tfrac{i}{2}X_\alpha{}^{\gamma}\epsilon_{\gamma\delta}}{m_\Psi^4}+\dots\right)\left(iD^{\dot{\beta}\alpha}\frac{\partial \mathcal{L}_\Psi^\text{int}}{\partial \eta^{\dagger\dot{\beta}}}+m_\Psi\frac{\partial\mathcal{L}_\Psi^\text{int}}{\partial\chi_\alpha}\right)+\text{h.c.}\\
		&=\frac{\partial \mathcal{L}_\Psi^{\text{int}}}{\partial \bar{\Psi}}\lr{\frac{1}{m_\Psi^2}+\frac{D^2+\tfrac{1}{2}X_{\mu\nu}\sigma^{\mu\nu}}{m_\Psi^4}+\dots}\left(i\slashed D+m_\Psi\right)\frac{\partial \mathcal{L}_\Psi^{\text{int}}}{\partial \Psi}\, ,
	\end{aligned}
\end{equation}
where we have used
\begin{equation}
	\sigma^{\mu\nu}\equiv\frac{i}{2}\left[\gamma^\mu,\gamma^\nu\right].
\end{equation}
For Majorana fields we can use the replacements $\eta^\alpha\,\to\,\chi^\alpha$ and $m_\Psi\,\to\, \tfrac{1}{2}m_\Psi$\, .

\subsection{Vector field}\label{sec:cdev}
A heavy vector field $\mathbf{V}^\mu$ leads to
\begin{equation}
	\label{eq:procalag}
	\mathcal{L}_{\text{S}}\supset -\frac{1}{4}\lr{D_\mu \mathbf{V}_\nu^{*}-D_\nu \mathbf{V}_{\mu}^*}\lr{D^\mu \mathbf{V}^\nu-D^\nu \mathbf{V}^\mu}+\frac{1}{2}m_{\mathbf{V}}^2\mathbf{V}^{*}_{\mu} \mathbf{V}^\mu-\frac{\partial \mathcal{L}_{\mathbf{V}}^{\text{int}}}{\partial \mathbf{V}^{*}_{\mu}} \mathbf{V}^{*}_{\mu}\, ,
\end{equation}
where we have used the field strength $X^{\mu\nu}=-i\left[D^\mu,\, D^\nu\right]$. The EOM can be obtained as
\begin{equation}
	D^2 \mathbf{V}^\mu -D^\nu D_\mu \mathbf{V}^\nu + m_{\mathbf{V}}^2\mathbf{V}^\mu-\frac{\partial \mathcal{L}_{\mathbf{V}}^{\text{int}}}{\partial \mathbf{V}^{*}_{\mu}}=0\, ,
\end{equation}
which we solve with the classical vector field
\begin{equation}
	\begin{aligned}
		(\mathbf{V}_{\text{cl}})_{\rho}&=\left(D^2\eta^{\mu\rho} -D^{\nu}D^{\mu}\eta_{\nu\rho}+ m_{\mathbf{V}}^2\eta^{\mu\rho}\right)^{-1}\frac{\partial \mathcal{L}_{\mathbf{V}}^{\text{int}}}{\partial  \mathbf{V}_\mu^*}\\
		&=\frac{1}{m_{\mathbf{V}}^2}\left(\eta_{\mu\rho}+\frac{D^2\eta_{\mu\rho} -D_{\nu}D_{\mu}\eta^{\nu\rho}}{m_{\mathbf{V}}^2}+\dots\right)\frac{\partial \mathcal{L}_{\mathbf{V}}^{\text{int}}}{\partial \mathbf{V}^*_\mu}\, .
	\end{aligned}
\end{equation}
This solution then leads to a series of interaction terms without the heavy vector field
\begin{equation}
	\begin{aligned}
		\mathcal{L}_{\text{eff}}\supset \mathcal{L}_{\text{eff}}^{\mathbf{V}} &= \frac{\partial \mathcal{L}_{\mathbf{V}}^{\text{int}}}{\partial \mathbf{V}_\nu}\lr{\frac{\eta_{\mu\nu}}{m_{\mathbf{V}}^2}+\frac{D^2\eta_{\mu\nu} -D_{\nu}D_{\mu}}{m_{\mathbf{V}}^4}+\dots}\frac{\partial \mathcal{L}_{\mathbf{V}}^{\text{int}}}{\partial \mathbf{V}_\mu^*}\\
	\end{aligned}
\end{equation}
%%%%%%%%%%%%%%%%%%%%%%%%%%%%%%%%%%%%%%%%%%%%%%%%%%%%%%%%%%%%%%%%%%%%%%%%%%%%%%%%%

\bibliographystyle{JHEP}
\bibliography{References}
\end{document}